\documentclass[twocolumn]{aastex631}
\usepackage{array}
\usepackage{amsmath}
\usepackage{makecell}

\usepackage{multirow}
\usepackage{nicefrac}

\shorttitle{Halo asymmetry and galaxy clustering}

\begin{document}
\title{Halo asymmetry in the modelling of galaxy clustering}

\author[0000-0002-3818-8315]{Anna Durkalec}
\affiliation{National Centre for Nuclear Research, ul. Ludwika Pasteura 7, 02-093 Warszawa, Poland}

\author{Agnieszka Pollo}
\affiliation{National Centre for Nuclear Research, ul. Ludwika Pasteura 7, 02-093 Warszawa, Poland}
\affiliation{Astronomical Observatory of the Jagiellonian University, Orla 171, 30-001 Cracow, Poland}

\author{Ummi Abbas}
\affiliation{INAF - Osservatorio Astrofisico di Torino, Via Osservatorio 20, Pino Torinese, Italy 10025}

\begin{abstract}
Conventional studies of galaxy clustering within the framework of halo models typically assume that the density profile of all dark matter haloes can be approximated by the Navarro-Frenk-White (NFW) spherically symmetric profile. However, both modern N-body simulations and observational data suggest that most haloes are either oblate or prolate, and almost never spherical.
In this paper we present a modified model of the galaxy correlation function.
In addition to the five ``classical'' HOD parameters proposed by \cite{Zheng2007}, it includes an additional free parameter $\phi$ in the modified NFW density profile describing the asymmetry of the host dark matter halo.

Using a subhalo abundance matching model (SHAM), we populate galaxies within BolshoiP N-body simulations. 
We compute the projected two-point correlation function $w_p(r_p)$ for six stellar mass volume limited galaxy samples.
We fit our model to the results, and then compare the best-fit asymmetry parameter $\phi$ (and other halo parameters) to the asymmetry of dark matter haloes measured directly from the simulations and find that they agree within 1$\sigma$.
We then fit our model to the $w_p(r_p)$ results from \cite{Zehavi2011} and compare halo parameters. 
We show that our model accurately retrieves the halo asymmetry and other halo parameters.
Additionally, we find $2-6\%$ differences between the halo masses ($\log M_{min}$ and $\log M_1$) estimated by our model and ``classical'' HOD models.
The model proposed in this paper can serve as an alternative to multiparameter HOD models, since it can be used for relatively small samples of galaxies. 

\end{abstract}

\section{Introduction}
In the Standard Cosmological Model ($\Lambda$CDM), the baryonic components of galaxies, in the form of stars, gas and dust, are thought to be embedded in dark matter (DM) haloes. 
This dark matter component is dominant and, as such, drives the evolution of the large-scale structure of the Universe.
In this scenario, the growth of structure is thought to be hierarchical.
Dark matter overdensities first collapsed into small haloes, which then grew progressively over time both through the steady gravitational inflow of surrounding dark matter and through halo mergers.
The baryonic component accreted at the centres of these haloes indirectly followed the evolution of the dark matter structure, forming the complex structures observed in the local Universe \citep[e.g., ][]{White1978, Kaiser1984, Bardeen1986, Mo1996, Kauffmann1997}.

Today, direct observations of dark matter are impossible, except for those provided by weak lensing studies \citep{Hoekstra2013, Mandelbaum2014, Mandelbaum2018}.
Therefore, methods that use a combination of galaxy observations and theoretical models that describe the relationship between galaxies and dark matter have been developed to study the properties of dark matter halos.
On large cosmic scales, one of the most widely used methods are Halo Occupation Distribution models \citep[HOD, e.g.,][]{Seljak2000, Peacock2000, Magliocchetti2003, Zehavi2004, Zheng2005}, which are able to constrain the properties of the DM haloes by modelling the clustering properties of the galaxies that reside within them.

Within the framework of empirical halo modelling, it is common practice to assume a spherical symmetry of DM haloes, with halo density profiles following the form proposed by \cite{Navarro1997} (often referred to as the NFW profile).
However, both N-body numerical simulations and observations suggest that these assumptions, while a good first approximation, may not reflect the true shape and mass distribution of dark matter haloes.
As a result, most studies lack information about the true shape of dark matter haloes and their asymmetries. The assumption of spherical symmetry could also influence the measurements of the halo masses and thus the final conclusions of such studies. Indeed, it has been shown in weak-lensing studies that halo traxiality is the main source of uncertainty in halo mass estimates \citep{Osato2018, McClintock2019, Zhang2022}.  

Moreover, spherically symmetric haloes are largely ruled out by N-body numerical simulations, which show that a typical DM halo is a triaxial spheroid (often asymmetric), which tends to be prolate in shape \citep[e.g.,][]{Frenk1988, Dubinski1991, Warren1992, Cole1996}. 
In addition, the specific shape of a DM halo depends strongly on both its mass and redshift, with more massive (or/and high redshift) haloes being less spherical and more prolate than the less massive and/or lower redshift ones \citep{Jing2002,Hopkins2005,Kasun2005,Allgood2006,Bett2007,Maccio2007,Munoz2011,Schneider2012,Vega2017}.
Deviations from the spherical shape in the most massive haloes, such as those hosting clusters of galaxies, are mostly due to their frequent mergers with smaller/less massive haloes, usually from one direction (i.e. along filaments), which prevents the DM halo from maintaining a relaxed ellipsoidal shape \citep[e.g.,][]{lokas2000}.

The results of N-body simulations are also strongly supported by observational evidence, mostly from strong and weak gravitational lensing observations. Most of these observations suggest that either oblate or prolate spheroidal shapes of dark matter haloes are the most common, especially for the most massive, cluster-sized DM haloes \citep[e.g.,][]{Carter1980, Sackett1990, Evans2009, Kawahara2010,Sayers2011,Oguri2010,Oguri2012,Despali2017, Chiu2018,Okabe2020,Lau2021, Hellwing2021, Gonzalez2022}.
Measurements based on the proper motions of globular clusters in our own Milky Way from the Gaia data \citep{Gaia2018} also favour a prolate rather than a spherical shape of the Milky Way dark matter halo \citep{Posti2019, Watkins2019}.

Simplified assumptions about the symmetry and mass profile can then lead to an under/overestimation of the size of the DM halo and an erroneous estimate of the number of galaxies that may reside in it.
\cite{Hayashi2012} found that mass estimates of dark matter dominated dwarf spheroidal galaxies are sensitively dependent on the assumed mass profiles and shapes of the dark matter halos.
They showed that the $M_{300}$ mass (mass enclosed in a spheroid with a major axis length of 300 pc) can be significantly overestimated when spherical symmetry is assumed for both stellar and dark halo density profiles.
On larger scales, \cite{Corless2007} investigated the effect of the assumption of spherical symmetry of the DM halo on measurements of massive clusters based on weak gravitational lensing and found that halo masses can be overestimated by up to $50\%$ if halo asymmetry is not taken into account.

Independent of the problem of the shape of DM haloes
numerous studies based on hydrodynamical simulations and semi-analytical models have shown that galaxy occupation is strongly related to many other secondary halo properties \citep[e.g.][]{Artale2018, Zehavi2018, Hadzhiyska2020, Xu2021, Yuan2021}. These effects are commonly referred to as Galaxy Assembly Bias, and have been shown to be a significant source of error in studies of the galaxy-halo relationship \citep[e.g.][]{Zentner2014}.

In recent years, many studies have focused on improving and extending HOD models to account for these effects. On the empirical modelling side, \cite{Hearin2016} created so-called decorated HOD models that minimally expand the parameter space with respect to ``classical'' HODs to account for the assembly bias. Other groups took advantage of the precision and volume of the latest cosmological simulations and created HOD frameworks that include multiple parameters to account for additional effects such as velocity bias, environment based bias, and concentration bias \citep[e.g.,][]{Zheng2016, Wibking2019, Wibking2020, Zhai2019, Yuan2022}.

However, a ``classical'' HOD model coupled with the NFW density profile is still widely used in studies of galaxy clustering, despite its known weaknesses and in the presence of superior models \citep[e.g. see recent work by][]{Gao2022, Lange2022, Linke2022, Yung2022, Qin2022, Zhai2022, Harikane2022, Alonso2023, Petter2023}. This shows the need for simple, easy-to-apply and computationally cheap HOD models that describe observed galaxy clustering reasonably well.

In this paper we take such a minimalist approach. We present an empirical 2-point correlation function model that consists of the combination of the ``classical'' HOD model and a modified NFW density profile that includes an additional free parameter. This parameter accounts for the DM halo asymmetry. Our main goal is to show that a simple modification of the NFW density profile is sufficient to make a relatively unbiased prediction of the galaxy-halo connection, especially when measurements are made on relatively small data samples. We focus mainly on the modelling of the projected correlation function $w_p(r_p)$, as this is the most commonly used statistic to measure galaxy clustering based on observations. To be clear, our model is certainly not designed to be used to populate large N-body simulations, as it does not take into account any of the secondary effects mentioned above. Instead, we focus on its applicability to clustering studies on small samples of galaxies, for which extended multi-parameter HODs prove to be overly complex.

We begin by fitting the projected two-point correlation function $w_p(r_p)$ of the mock galaxy catalogue with our model. Using the results of this fitting, we show the accuracy of this new model and its usability for clustering measurements. We then apply our model to correlation functions measured by \cite{Zehavi2011} and show the differences between results obtained with and without the dark matter halo shape assumption. 

The paper is structured as follows. 
In section \ref{sec:2_cosm_sim} we briefly describe the properties of the cosmological simulations used in this work, the methods used to identify DM haloes and to populate them with galaxies. We then examine the main properties of dark matter haloes relevant to this work, and finally present the selection of the galaxy samples. 
Next, in section \ref{sec:3_DMhalo_shape} we present the formulae we use to describe the DM halo shape and introduce a free parameter $\phi$ which is then included in the modified NFW density profile to account for the DM halo asymmetry. 
The methods for measuring and modelling the galaxy correlation function are presented in section \ref{sec:4_CFandHODmodel}. 
Our results are presented in section \ref{sec:5_results}. There we also present the differences between the halo masses estimated using our model and the results based on the HOD model assuming halo spherical symmetry from \cite{Zehavi2011}. The limitations and applicability of our model are discussed in section \ref{sec:discussion}. Finally, a summary and conclusions are presented in section \ref{sec:6_summary}.

Throughout this paper, we refer to distances in comoving units. We adopt a flat $\Lambda$CDM cosmology, with $\Omega_M = 0.307$, $\Omega_{\Lambda}=0.693$ \citep{Planck2014} for measurements based on the mock galaxy catalogue, and $\Omega_M = 0.3$, $\Omega_{\Lambda}=0.7$ for measurements based on the correlation functions from \cite{Zehavi2011}. In both cases the distances are given in units of h$^{-1}$ Mpc (where $h = H_0/100$ km s$^{-1}$ Mpc$^{-1}$).
%

\section{Data}
\label{sec:2_cosm_sim}
Cosmological N-body simulations are proving to be a unique tool for detailed studies of large-scale structure evolution, especially within the $\Lambda$CDM model framework. 
This is also true in the context of our work. 
The very good resolution of modern N-body simulations allows detailed studies of dark matter structures and their asymmetries even on small cosmic scales ($<1$ Mpc). 

In this section we briefly describe the main features of the large N-body simulation used in our work - Bolshoi- Planck (BolshoiP).
For a detailed description of this data set, we refer the reader to the dedicated paper by \cite{Klypin2016}.
We also present the methods used to populate the galaxies within the simulated DM haloes, and the selection of stellar mass subsamples from these galaxy sets. 
%

\subsection{Cosmological N-body simulations}
In this paper we use the Bolshoi-Planck N-body dark matter-only cosmological simulation \citep[BolshoiP,][]{Klypin2011, Klypin2016}.
The parameters describing this simulation, such as comoving volume, number of particles and mass resolution, are listed in Table \ref{tab:simulations_parameters}. 
Below we briefly describe only some of its aspects that are important in the context of our work. 
%

%
\begin{deluxetable}{ccc}
    \tablecaption{Main properties of BolshoiP cosmological simulations used in this work. \label{tab:simulations_parameters}}
    \tablenum{1}
    \tablehead{\colhead{Comoving volume } & \colhead{Number of} & \colhead{Mass resolution } \\
    \colhead{(Mpc$^3/h^3$)} & \colhead{particles} & \colhead{$(\times10^9 M_{\odot}/h)$}}
     \startdata
     250$^3$	& 2048$^3$	&	$0.15$	\\
     \enddata
\end{deluxetable}

The BolshoiP is the high-resolution simulation of $2048^3$ ($\approx8.6\times10^9$) collisionless dark matter particles distributed within a comoving volume of $(250$ $h^{-1}$ Mpc $)^3$ over the redshift range from $z=80$ to present day. In this paper we focus on the redshift $z=0$. 
BolshoiP assumes a flat $\Lambda$CDM cosmology, with cosmological parameters obtained from the Planck mission data and published by \cite{Planck2014} (i.e., $\Omega_M = 0.307$, $\Omega_{\Lambda}=0.693$, $h=0.7$, $n=0.96$, $\sigma_8=0.82$).
The high mass resolution, relatively large volume box, and updated cosmological parameters make this simulation ideal for clustering studies such as those presented in this paper.
%

\subsection{Creating the galaxy mock catalogue}
\label{sec:sample_creation}
We used the publicly available Python package Halotools \citep[v0.7, see][for a description of the first release]{Hearin2017} to identify the DM haloes in the BolshoiP simulations. This package includes the ROCKSTAR halo finder algorithm \citep{Behroozi2013}, which uses the precise approximation of the DM halos by associating non-spherically symmetric ellipsoids with their mass distribution \citep{Allgood2006}.
We focus on DM haloes with masses in the range of $10^{10.5} M_{\odot}$ to $10^{15} M_{\odot}$ at $z=0$.

In the next step we populated the haloes and subhaloes with galaxies - creating mock catalogues. We use a subhalo abundance matching technique \citep[SHAM,][]{Kravtsov2004}.
To implement this method we again used the Halotools python package and one of its pre-built galaxy-halo connecting model: the stellar-to-(sub)halo (SMHM) models proposed by \cite{Behroozi2010}.
This model connects the stellar mass of galaxies to the mass of their host DM haloes and subhaloes by assuming a direct relation given in parametrized form of the SMHM function by \cite{Behroozi2010} (see Table 2 therein). The level of intrinsic scatter is set to 0.2 dex.
The central galaxies are placed at the centre of their host DM halo and the satellite galaxies at the centres of subhaloes. 
The most important aspect of this model is that the true distribution of subhaloes is not assumed to be spherically symmetric, so the distribution of satellite galaxies reflects the asymmetries of the host DM halo. These mock catalogues have been created without any additional observational strategies. 

We then further divide galaxy mock catalogue into six stellar-mass volume limited subsamples; M1: $\log(M_{\ast}/M_{\odot})>9.5$, M2: $\log(M_{\ast}/M_{\odot})>10.0$, M3: $\log(M_{\ast}/M_{\odot})>10.5$, M4: $\log(M_{\ast}/M_{\odot})>10.75$,   M5: $\log(M_{\ast}/M_{\odot})>11.0$, and M6: $\log(M_{\ast}/M_{\odot})>11.25$. 
The general properties of these subsamples, such as a number of galaxies, mean stellar mass, and mean host DM halo asymmetry parameters: the triaxiallity parameter $T$ (see Section \ref{subsec:T_halo_asymmetry}) and proposed in this work parameter $\phi$ (described in Section \ref{subsec:deviation_phi}) are presented in Table \ref{tab:subsamples_sm}.
%

\begin{deluxetable}{cccccc}
    \tablecaption{Number of galaxies $N_{gal}$, median stellar mass $\log M_{\ast}^{med}$, median host halo triaxiality parameter $T_{med}$ and median host halo asymmetry parameter $\phi_{med}$ for six stellar mass selected subsamples used in this work. All stellar masses are given in $(M_{\odot})$. \label{tab:subsamples_sm}}
    \tablenum{2}
    \tablehead{
    \colhead{Sample} & \colhead{$\log M_{\ast}^{min}$} & \colhead{$N_{gal}$} & \colhead{$\log M_{\ast}^{med}$} &	\colhead{$T_{med}$}	& \colhead{$\phi_{med}$}} 
    \startdata
    M1  &   $9.50$       &   $373,820$   &   $10.01$       & $0.65$  &   $0.75$ \\
	M2  &   $10.00$      &   $189,475$   &   $10.29$       & $0.67$  &   $0.72$ \\
	M3	&	$10.50$		&   $46,820$    & 	$10.66$		  &  $0.71$		& $0.68$   \\ 
	M4  &   $10.75$     &   $14,342$	&	$10.87$		  &  $0.73$		& $0.64$  \\ 
	M5	&   $11.00$		& 	$2,974$	    &   $11.09$		  &	 $0.75$		& $0.61$ \\ 
	M6  &   $11.25$		&	$381$		&	$11.31$		  &  $0.78$		& $0.57$
    \enddata
\end{deluxetable}

\section{Dark matter halo shape}
\label{sec:3_DMhalo_shape}
In this section we define the asymmetry parameter $\phi$. 
To better explain how this parameter relates to the halo shape, we first show its relation to the triaxiality parameter $T$. 
We then show how it can be used as a free parameter to modify the NFW density profile later used in the modelling of the galaxy correlation function.
%

\subsection{Standard description of halo asymmetry- traxiality $T$}
\label{subsec:T_halo_asymmetry}
The shape of a halo is usually characterised by three ellipsoidal axes {\it a, b, c}, with $a\geqslant b \geqslant c$, usually expressed as the ratio of the second to the largest axis to the largest axis (${\bf{\mu}}={b}/{a}$) and the ratio of the smallest axis to the largest axis (${\bf{\eta}}=c/a$). 
The triaxiality parameter $T$ is a combination of these two parameters \citep{Franx1991}:  
\begin{equation}
T=\frac{1-\bf{\mu}^2}{ 1 - \bf{\eta}^2}
\end{equation}
Three shapes of DM halos can be defined based on the value of the parameter $T$: oblate when $0<T<1/3$, triaxial when $1/3<T<2/3$ and prolate when $2/3<T<1$.

The value of the parameter $T$ depends strongly on the mass of the DM halo, with more massive haloes being more asymmetric and their overall halo shape often being prolate \citep[e.g.,][]{Vega2017}.
This is also the case in our halo catalogues. 
The prolate halo shape dominates and accounts for more than 50$\%$ of all possible halos in our sample. 

\begin{figure}
	\centering
		\includegraphics[scale=0.93]{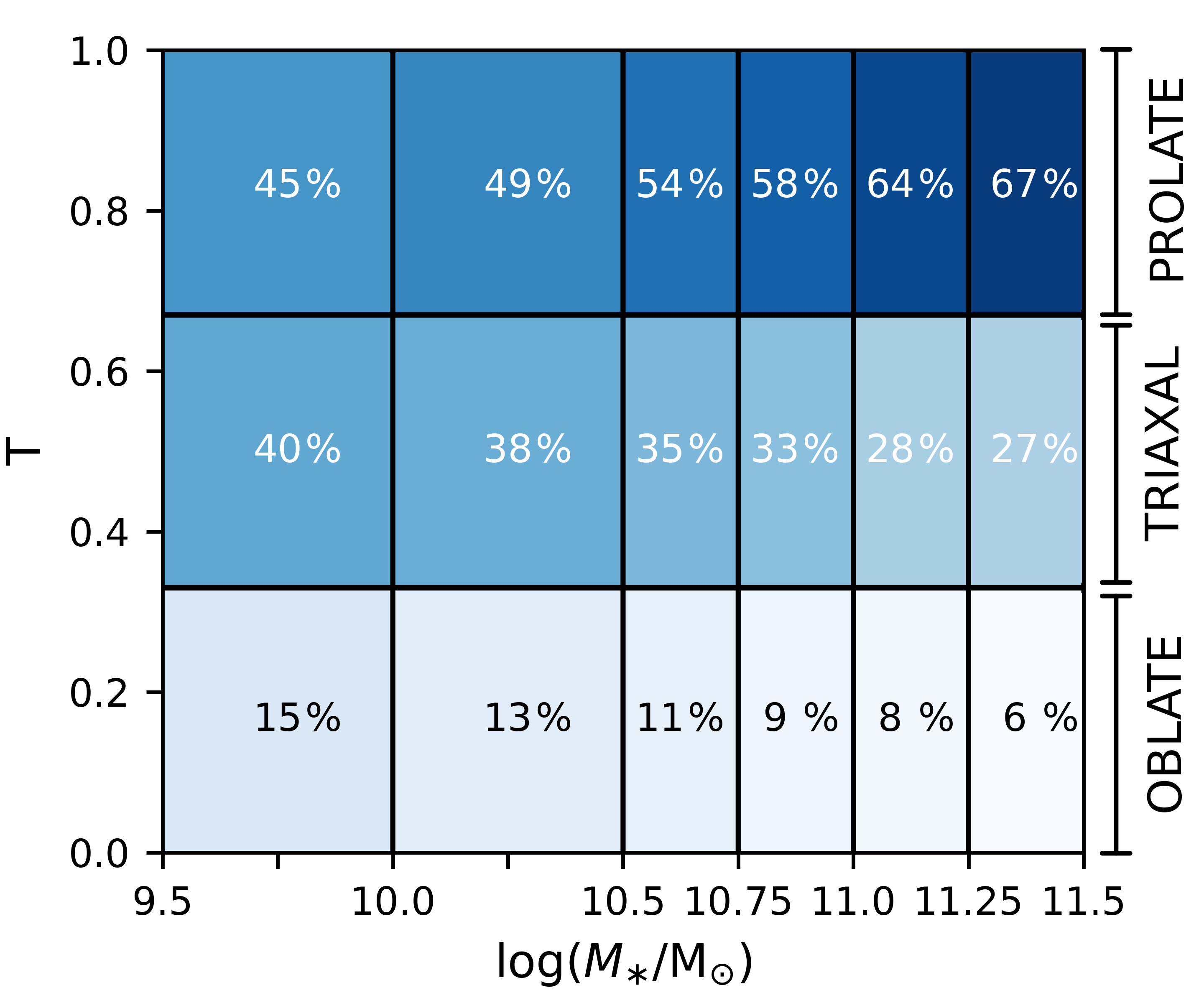} 	
	\caption{Triaxiality parameter $T$ of the host DM halo as a function of the stellar mass of the galaxies residing in this halo. Results for mock galaxy catalogues populated in BolshoiP simulations.
	The percentage of galaxies residing in a halo with one of the three types of asymmetry varies with stellar mass. The most massive galaxies are most likely to be in the prolate halos. This means that the halo shape might be the important factor in galaxy clustering modelling of these galaxies.
	}
	\label{fig:T_lum_sm_both}
\end{figure}

It is well established that the halo mass is strongly correlated with the stellar mass of the galaxy \citep[see e.g.][]{Meneux2008, Marulli2013, Beutler2013, Dolley2014, Skibba2015, Durkalec2018}.
Since the most massive haloes are also the most asymmetric, it is not surprising that in our mock galaxy catalogues we found a strong correlation between the host halo asymmetry parameter $T$ and the stellar mass of the hosted galaxies.
As shown in Figure \ref{fig:T_lum_sm_both}, about 67$\%$ of the most massive galaxies, with stellar masses greater than $10^{11.25} M_{\odot}$, reside in halos of asymmetric, prolate shape.
This percentage drops to 45$\%$ for galaxies with stellar masses less than $10^{10} M_{\odot}$. This means that the introduction of halo shape dependence in halo modelling of galaxy clustering will be particularly important for high mass samples.
%

%

%

%

\begin{figure}
    \centering
	\includegraphics[scale=0.93]{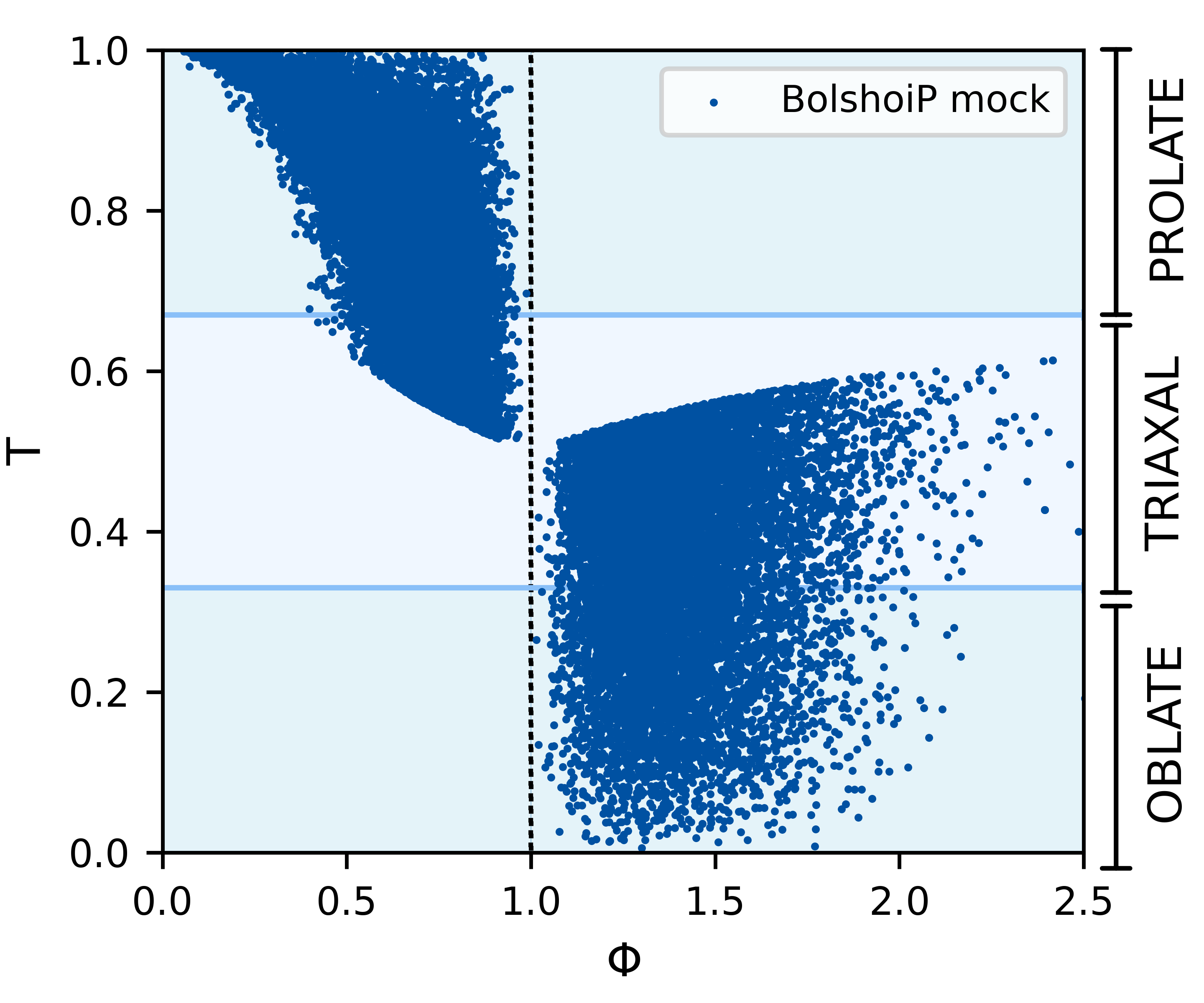}
	\caption{Parameter T as a function of $\phi$ measured for host DM halos.
		Each point represents a single host DM halo.
			$\phi = 1$ indicates a spherically symmetric halo.
		For host DM haloes with $T > 0.5$ (indicating a more prolate type of shape asymmetry) the $\sim 95\%$ of the sample haloes take values of $\phi < 1$. 
		For $T < 0.5$ (i.e. DM haloes with oblate shape asymmetry) the parameter is $\phi > 1$.}
	\label{fig:phi_T}
\end{figure}
%

\subsection{Defining a novel asymmetry parameter $\phi$}
\label{subsec:deviation_phi}
Introducing the common description of halo asymmetry - the traxiality $T$ - into the halo model would require us to use two additional free parameters ($\mu$ and $\eta$).
This could be computationally demanding for many studies.
As an alternative, we propose a new asymmetry parameter $\phi$. 
It has been constructed to measure how much the shape of the DM halo deviates from spherical symmetry.
In our definition of $\phi$ we assume spherical symmetry along one axis and measure the deviation from this symmetry along the other axis.
This approach greatly simplifies the implementation of halo asymmetry in halo models. The asymmetry parameter $\phi$ is defined as follows:
\begin{equation}
\begin{split}
\rm{if} \ \ d = A : &  \phi = \frac{a+b}{2c}  \\
\rm{if} \ \ d = B : &  \phi = \frac{b+c}{2a},  \\ 
\end{split}
\label{eq:phi}
\end{equation}
where $A = |a - b|$, $B = |b - c|$, $d = \min(A,B)$, and $a$, $b$, $c$ are the three axes defining the ellipsoidal shape of the DM halo ($a \geqslant b \geqslant c$). 

These two parameters - the traxiallity parameter and the newly defined asymmetry parameter $\phi$ - are obviously not independent. 
Their relationship is shown in Figure \ref{fig:phi_T}. 
There are two distinct areas in the plot - regions with $\phi$ less than one and regions with $\phi$ greater than one. 
The region of $\phi < 1$ correlates with values of $T$ between $0.5$ and $1$, indicating that host DM haloes have mostly prolate shapes. 
On the other hand, $\phi > 1$ are correlated with values of $T$ between $0$ and $0.5$, i.e. host DM haloes are predominantly oblate in shape.

By construction, the parameter $\phi$ assumes a value of $\phi = 1$ for an idealised case of spherically symmetric haloes (i.e. when the ellipsoidal axes are equal to each other).       
The further away from this value, the greater the deviation from spherical symmetry (i.e. the greater the asymmetry of the DM halo). $\phi > 1$ defines halos with oblate shape asymmetry, while $\phi < 1$ defines a prolate shape.

\begin{figure}[htb!]
\centering
	\includegraphics[scale=0.93]{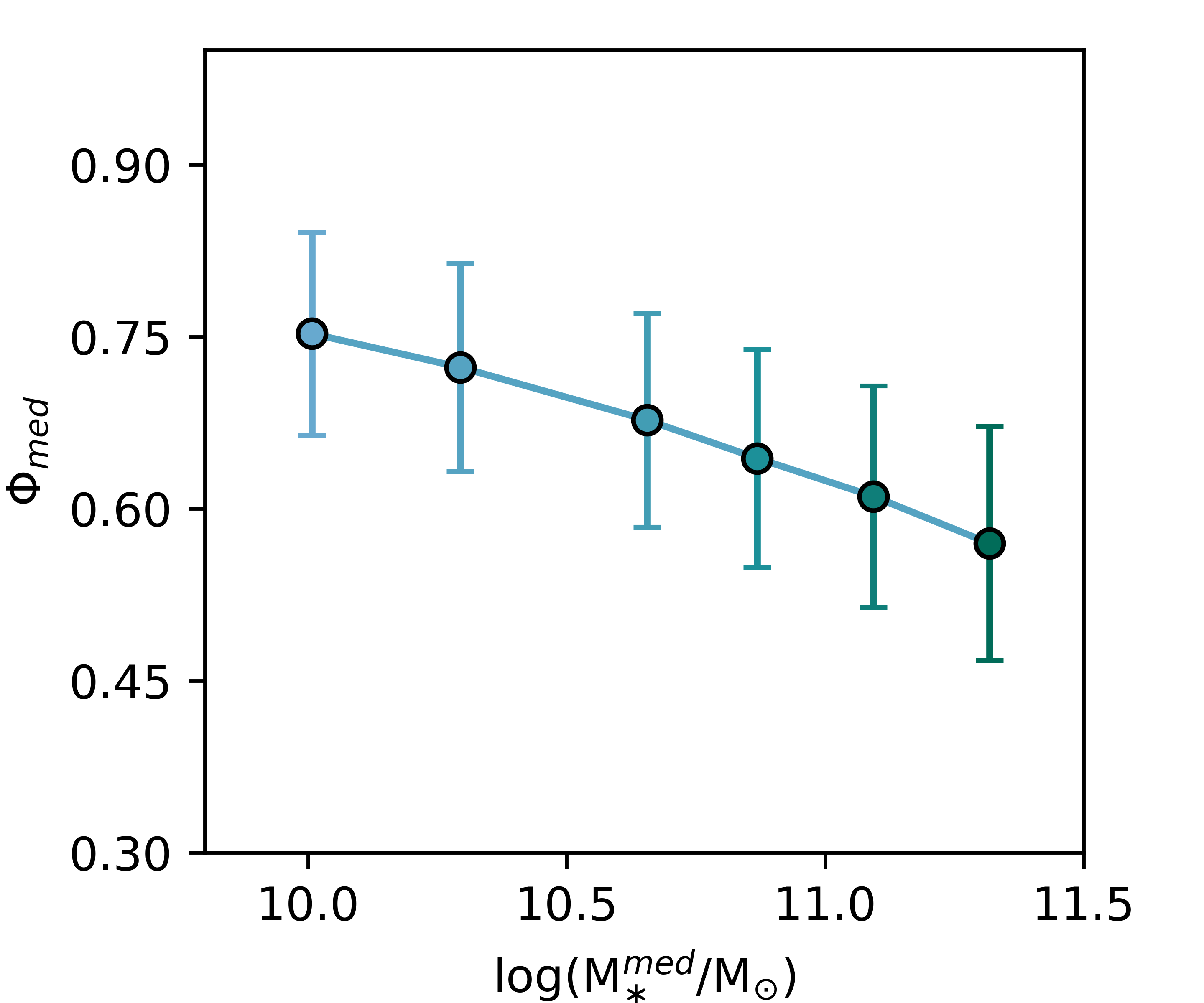}
	\caption{Median value of the host halo asymmetry parameter $\phi$ as a function of a median stellar mass of galaxy subsamples selected from the BolshoiP mock galaxy sample. The value of $\phi$ is obtained using the ellipsoid axes a, b and c available in the halo catalogue. Error bars represent the standard deviation. As expected, the asymmetry of the host halo increases with stellar mass and is strongest for the most massive sample.}
	\label{fig:phi_sm}
\end{figure}

As an example of how this parameter should be interpreted, we use the mock galaxy catalogue. Since the majority of mock galaxies are located within prolate haloes and most massive galaxies are more likely to be in the most asymmetrical haloes (see Figure \ref{fig:T_lum_sm_both}), we expect $\phi<1$ for all galaxy stellar mass samples, and its value to decrease with increasing stellar mass. Indeed, as shown in Figure \ref{fig:phi_sm}, $\phi$ is less than one for all samples, and the median, $\phi_{med}$, systematically moves away from $\phi=1$ (marking the spherically symmetric halo) with increasing stellar mass of the hosted galaxies. From $\phi_{med} = 0.75$ for the least massive galaxies with mean stellar masses $10^{10.01} M_{\odot}$ (sample M1), to $\phi_{med} = 0.57$ for the most massive galaxies with median stellar masses of $10^{11.31} M_{\odot}$ (sample M6). Similar results are therefore expected from the projected correlation function fit.
%

%

\subsection{Defining modified NFW density profile with asymmetry parameter $\phi$}
\label{sec:modified_dm_profile}
One of the most important components of the halo model is the density profile of the dark matter halo.
The most commonly used in the literature is the so-called Navarro-Frenk-White (NFW) density profile proposed by \cite{Navarro1997}. 
This symmetric profile is easy to adapt in to halo models, but it may not properly reflect the real dark matter halo density profile. In particular, it does not take into account the possibility that the DM halo may have a non-spherical shape, which is a common occurrence as shown in section \ref{sec:3_DMhalo_shape}.  
To account for this halo property, we propose to extend the standard NFW density profile by adding an additional parameter $\phi$ (see Eq. \ref{eq:phi}), which can be interpreted as a deviation of the halo shape from spherical symmetry. 

In the general case, the dark matter NFW profile is defined as \citep{Navarro1997}:
\begin{equation}
\frac{\rho(R)}{\rho_{crit}}=\frac{\delta_c}{\frac{R}{R_S}\left(1+\frac{R}{R_S}\right)^2}, 
\end{equation}
where $\rho_{crit}$ is a present ($z=0$) critical mass density, $\delta_c$ is the overdensity of the DM halo, $R_s$ is a characteristic DM halo radius, and {\bf R} is a 3-dimensional vector described by the three ellipsoidal axes $a$, $b$, and $c$ as follows: 
\begin{equation}
R=\sqrt{\frac{x^2}{a^2} + \frac{y^2}{b^2} + \frac{z^2}{c^2}}. 
\label{eq:R}
\end{equation}
When $a=b=c$, the NFW density profile becomes spherically symmetric.
However, if we assume that only one of two pairs of axes $a$ and $b$, or $c$ and $b$ is approximately equal, then we can rewrite this equation as:
\begin{equation}
\begin{split}
R=\frac{1}{a}\sqrt{r^2 + \frac{z^2}{(c/a)^2}} \ \ \rm{if} \ \ a=b \\
R=\frac{1}{b}\sqrt{r^2 + \frac{x^2}{(a/b)^2}} \ \ \rm{if} \ \ b=c,  \\
\end{split}
\end{equation}
where the 2-dimensional vector $x^2 + y^2$ or $y^2 + z^2$ is represented by {\bf r}.

If we also assume (in the first approximation), that  {\bf r}  is proportional to $z$, then:
\begin{equation}
\begin{split}
R  \approx \frac{r}{a} \sqrt{1 + \frac{1}{(c/a)^2}} \quad  
\rm{or} \quad
R  \approx \frac{r}{b} \sqrt{1 + \frac{1}{(a/b)^2}} 
\label{eq:R_r}
\end{split}
\end{equation}
where $c/a$ and $a/b$ measure the asymmetry of DM halo profile shape (in two dimensions) with respect to the spherical shape. 
Rather than using  $\frac{c}{a}$ and $\frac{a}{b}$ we can take the average length of the two axes divided by the third one.
Now equation (\ref{eq:R_r}) assumes the form: 
\begin{equation}
\begin{split}
R  \approx \frac{r}{a} \sqrt{1 + \left(\frac{a+b}{2c}\right)^2} \ \ \rm{if} \ \ a=b \\
R  \approx \frac{r}{b} \sqrt{1 + \left(\frac{b+c}{2a}\right)^2} \ \ \rm{if} \ \ b=c \\,
\label{fig:R_two}
\end{split}
\end{equation}
and the variable in the parenthesis can be rewritten making use of the parameter $\phi$ (as defined in Eqation \ref{eq:phi}). 
Finally, we can express $R$ as:
\begin{equation}
R  \approx \frac{r}{a\sqrt{2}}\sqrt{1 + {\phi^2}}\sim\frac{r}{b\sqrt{2}}\sqrt{1 + {\phi^2}}\ ,
\label{eq:R_final}
\end{equation}
where parameter $\phi$ has values in the range $ 0<\phi <\infty$, and a factor $\frac{1}{\sqrt{2}}$ is present due to the fact that $ \phi = 1 $ profile is required to be spherically symmetric.

\begin{figure}
	\centering
	\includegraphics[scale=0.93]{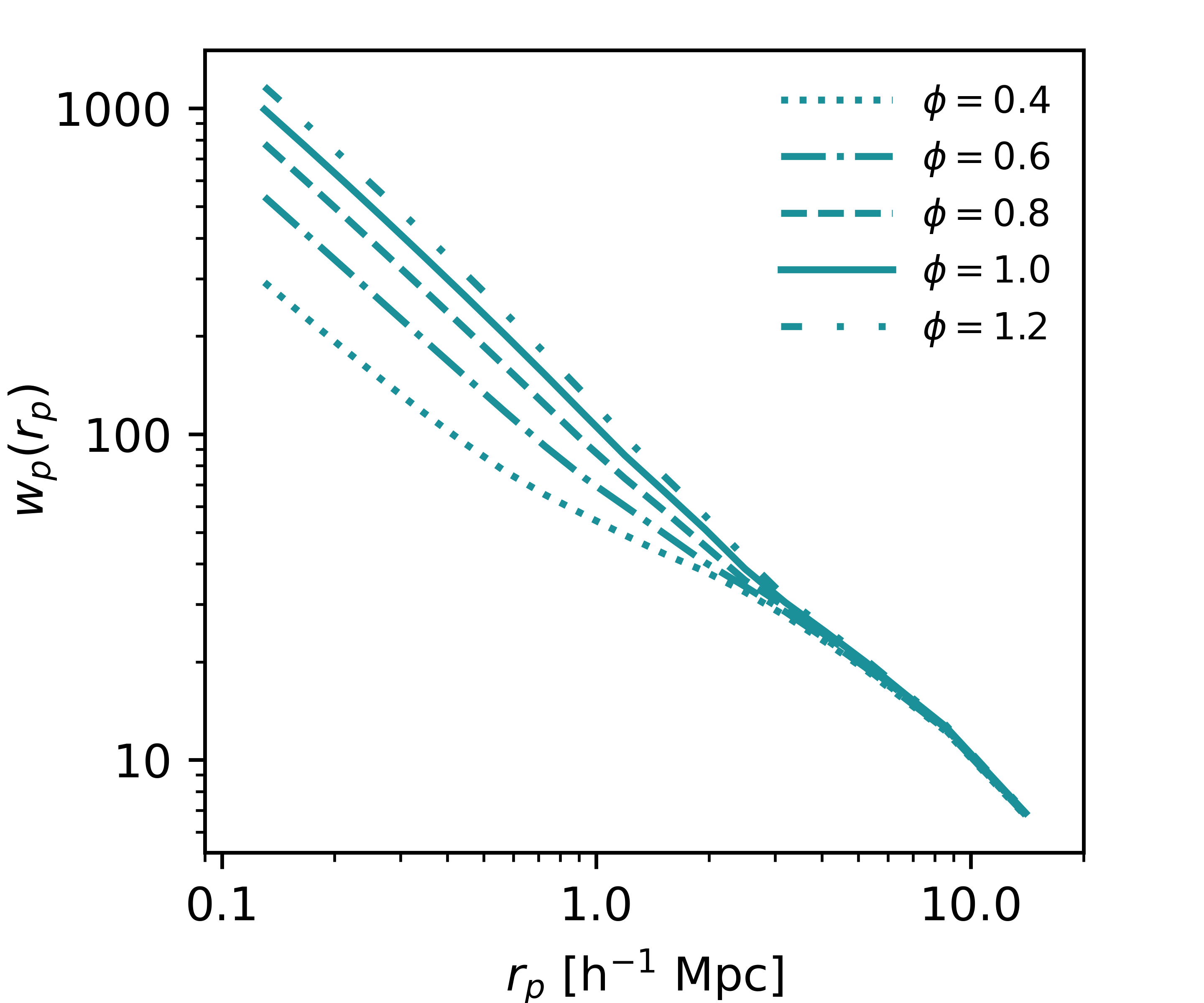}
	\caption{The modelled two-point correlation function for different halo asymmetries. Different lines mark the modelled correlation functions for different $\phi$ (as labelled) and other parameters fixed to the same value: $\log M_{min} = 12.40$, $\log M_1 = 13.86$, $\log M_0 = 10.58$, $\sigma_{logM} = 0.75$, $\alpha = 0.93$ (as results for sample M4, see Section \ref{sec:5_results}).}
	\label{fig:wprp_asymetry_test}
\end{figure}

The density profile modifications and the addition of the asymmetry parameter $\phi$ of course affect the modelled two-point correlation function.
In Figure \ref{fig:wprp_asymetry_test} we show how the modelled correlation function changes for different halo asymmetries while the other HOD parameters are fixed at the same values. 
As shown, the asymmetry of the halo mostly affects the correlations on scales of $<$1 h$^{-1}$ Mpc (one halo term), where it has a significant influence on the shape of the correlation function. 
The difference between a spherically symmetric halo (marked with $\phi = 1$) and an asymmetric (prolate) halo with $\phi=0.4$ at scales $r_p = 0.2$ h$^{-1}$ Mpc reaches $\Delta w_p(r_p) = 663.35$ (3.5 dex) in the case presented.
The assumption of spherical symmetry of the DM halo can therefore influence the results of the $w_p(r_p)$ fitting.

%
%

%
\section{Correlation function, the baseline HOD model and $w_p(r_p)$ fitting}
\label{sec:4_CFandHODmodel}

\subsection{Correlation function}
To measure the correlation functions presented in this paper \footnote{except those of \cite{Zehavi2011}} we used the Haltools v0.7 Python package \citep[v0.7, see][for a description of the first version]{Hearin2017}. 
For details we refer the reader to documentation of this code.
Here we will only present the most important details.

A commonly used estimator for the two-point correlation function was introduced by \citet{Landy1993}:
\begin{equation}
\xi(r_p,\pi)= \frac{N_R(N_R-1)GG}{N_G(N_G-1)RR} - 2\frac{(N_R-1)GR}{N_G RR} +1,
\end{equation}
where $N_G$ and $N_R$ are the total number of objects in the galaxy sample and random points generated in the same volume and with the same geometric properties as the real sample. 
$GG$, $GR$ and $RR$ are the number of galaxy-galaxy, galaxy-random and random-random pairs at a given separation radius. 
The random points catalogue has been constructed to contain 100 times more objects than the galaxy catalogue.
The projected 2-point correlation function is computed by integrating $\xi$ over $\pi$ between 0 and $\pi_{max} = 60$ Mpc/h:
\begin{equation}
w_p(r_p)=2\int_0^{\pi_{max}} \xi(r_p,\pi)d\pi.
\end{equation}
For each volume limited stellar mass galaxy sample this projected correlation has been measured in 12 equally spaced (on the logarithmic scale) radial bins, except for the most massive sample M6, where we used 8 bins due to the small number of galaxies available. In each case the correlation functions have been measured over the range from $r_p^{min} = 0.1$ h$^{-1}$ Mpc to $r_p^{max}= 16$ h$^{-1}$ Mpc.

The statistical errors of the correlation function measurements were estimated using a jackknife resampling method.
We created $N = 125$ equal-volume cubic boxes, each of sizes 50$^3$ Mpc$^3/$h$^3$ covering the entire volume of the galaxy sample. Then we created subsamples systematically leaving out one of the boxes.
The error covariance matrix, which describes a total dispersion between these samples, was computed using
\begin{equation}
C(w_{p,i},w_{p,j}) = \frac{N-1}{N} \sum_{k=1}^N(w_{p,i}^k - \bar{w}_{p,i})(w_{p,j}^k - \bar{w}_{p,j})
\label{eq:covariance}
\end{equation}
where $\bar{w}_{p,i}$ and $\bar{w}_{p,j}$ are the means of the correlation function in bins $i$ and $j$ respectively. 
%

%
\subsection{Halo Occupation Distribution} 
In our work we use the baseline five parameter HOD model described in \citep{Zheng2007}.
\begin{equation}
\begin{split}
\langle N_{cen} (M_h) \rangle & = \frac{1}{2}\left[ 1+ \rm{erf} \left( \frac{\log M_h - \log M_{min} }{\sigma_{\log M_h}} \right) \right], \\
\langle N_{sat} (M_h) \rangle & = \langle N_{cen} (M_h) \rangle \times \left(\frac{M_h-M_0}{M_1}\right)^\alpha.
\end{split}
\label{eq:HOD}
\end{equation}

This model has five free parameters: $M_{min}$, $M_1$, $M_0$, $\sigma_{\log M_h}$, and $\alpha$. 
$M_{min}$ denotes the minimum halo mass for which half of the DM halos contain a central galaxy above the adopted stellar mass (or luminosity) threshold for this sample.
$M_1$ denotes the satellite halo mass for which a DM halo contains on average one additional satellite galaxy, while $M_0$ denotes the cutoff mass scale.
The scatter between the stellar mass (or luminosity) of the galaxies and the halo mass is given by $\sigma_{\log M_h}$, while $\alpha$ is the power-law slope of the galaxy mean occupation function.

The core assumption of this HOD model is that the number of galaxies residing within the host DM halo is a function of the mass of that halo $\langle N_g(M_h) \rangle$ and that the total number of galaxies within an average halo is a sum of the average occupation of central $\langle N_{cen} (M_h) \rangle$ and satellite $\langle N_{sat} (M_h) \rangle$ galaxies:
\begin{equation}
\langle N_g(M_h) \rangle = \langle N_{cen} (M_h) \rangle +  \langle N_{sat} (M_h) \rangle.
\end{equation}

Finally, using the best-fit HOD parameters we are able to obtain the average DM halo mass $\langle M_h \rangle$ hosting a given galaxy population with
\begin{equation}
\langle M_h \rangle (z) = \int dM_h \ M_h \ n(M_h,z) \ \frac{\langle N_g \left(M_h \right) \rangle}{n_g\left(z\right)}
\label{eq:average masses}
\end{equation}
and the large scale galaxy bias $b_g$
\begin{equation}
b_g = \int dM_h b_h(M_h)n(M_h,z)\frac{\langle N_g(M_h) \rangle}{n_g(z)},
\label{eq:bias}
\end{equation}
where $n\left(M_h,z\right)$ is the DM mass function for which we adopted the fitting formula proposed by \cite{Tinker2008}, and $n_g\left(z\right)$ represents the number density of galaxies,
\begin{equation}
n_g(z) = \int dM_h \ n\left(M_h,z\right) \ \langle N_g\left(M_h\right)\rangle.
\end{equation}
%

\subsection{Model fitting procedure}
\label{sec:fitting}
To fit our model to the correlation functions, we use Markov Chain Monte Carlo (MCMC) methods. 
The MCMC sampling was done by implementing the affine-invariant ensemble sampler of \cite{Goodman2010}, which is provided by the publicly available Python library emcee \citep{Foreman2013}.
To generate the posterior parameters for each fit, we run an MCMC with $n = 25$ random walkers (chains), each of which explores the parameter space starting from different randomly chosen initial parameters. 
At each step of the random walk, a new set of HOD parameters is generated from a Gaussian distribution with a fixed variance $\sigma^2$ and is accepted if $\min[\exp(-(\chi^2_{new} - \chi^2_{old})), 1]$ is less than a random number generated from a uniform distribution in the range [0,1].
To compute $\chi^2$, we use the measured values of the projected correlation function $w_p(r_p)$ with the full error covariance matrix $\bf{C}$, and the number density of galaxies $n_g$ in each subsample as
\begin{equation}
\begin{split}
\chi^2= (w_p^{mod} - w_p^{obs})^T C^{-1} (w_p^{mod} - w_p^{obs}) \\
+ \left(\frac{(n_g^{mod} - n_g^{obs})}{\sigma_{n_g}}\right)^2
\end{split}
\end{equation}
where $w_p$ is a vector containing measurements of the two-point correlation function. 
We assume a $1\%$ uncertainty $\sigma_{n_g}$ for the observed number density. The suffixes "obs" and "mod" denote the values measured from galaxy mock (or observation) catalogues and HOD model predictions, respectively.  

The best-fit HOD parameters are determined by finding the 50th percentile of the marginal posterior probability distribution of all random walk realisations. Uncertainties are taken as the 16th and 84th percentiles.

For each fit, we use the following methods to ensure its convergence:
\begin{enumerate}
\item The total number of random walk steps for each chain realisation is determined by using the integrated                autocorrelation time $\tau$ \citep{Foreman2013}.
      Following this method, we first determine the $N_{burn}$ steps required for convergence. We assume 
      \begin{equation}
        N_{burn} > 50 \times \tau
       \end{equation}
      where $\tau$ is calculated using methods available in the emcee library.
      After reaching $N_{burn}$ we then continue for an additional number of steps not less than the size of $N_{burn}$ for the given galaxy sample. 
\item We ensure that the $\chi^2$ value for each random walk chain converges and stabilises at the lowest possible          value.
\end{enumerate}

\begin{figure}
\centering
	\includegraphics[scale = 0.93]{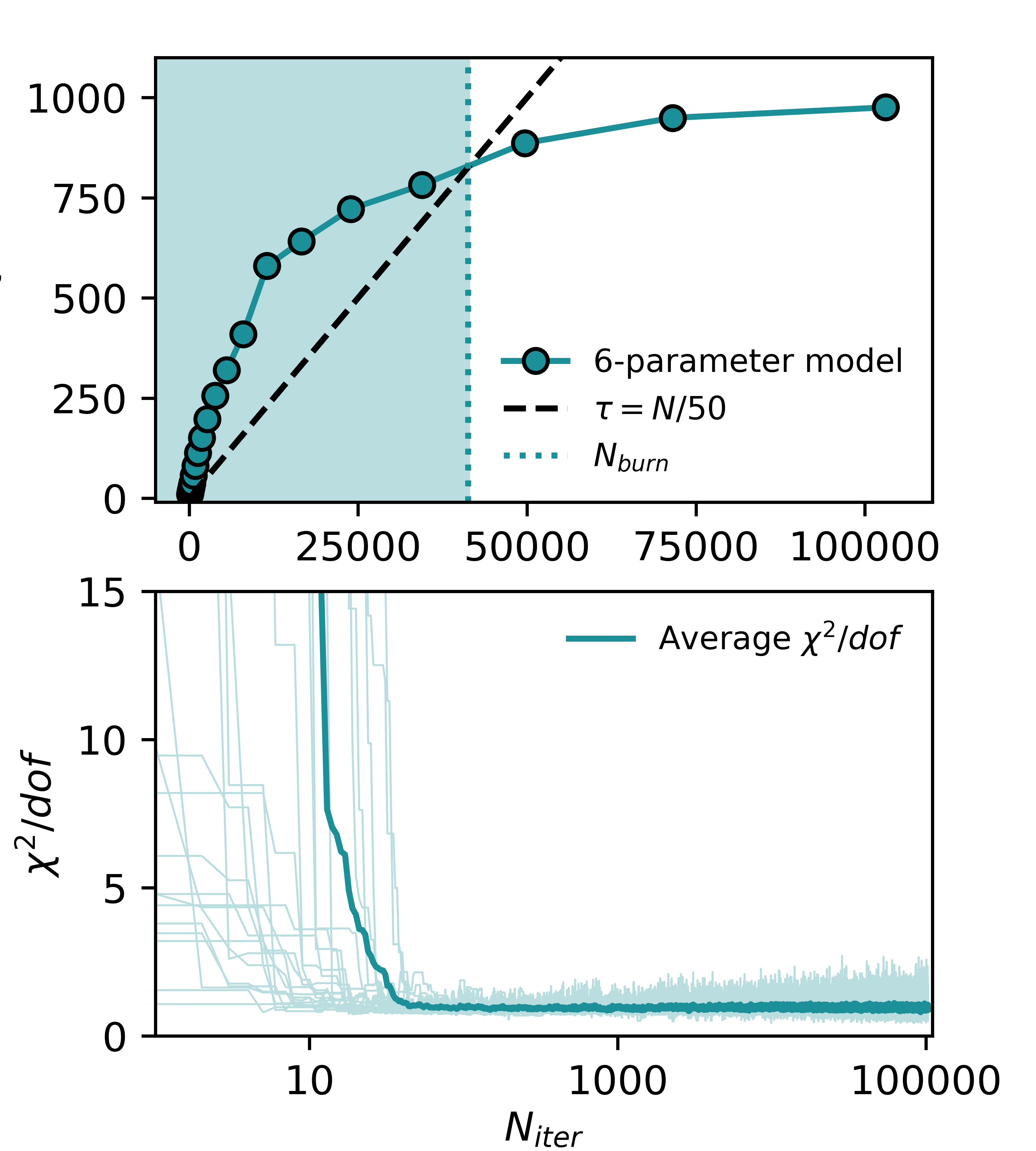}
	\caption{Top panel: Average integrated autocorrelation time $\tau$ as a function of the number of chain iterations for the correlation function modelling of mock galaxies from the M2 sub-sample. The filled circles mark the average autocorrelation time obtained for all six free parameters. The shaded area shows the $N_{burn}$ iterations necessary for the fit to converge, i.e. the number of iterations for which $\tau = N_{burn}/50$. Bottom panel: $\chi^2$/dof value as a function of iteration number for all 25 random walk chains.}
	\label{fig:tau and chi2}
\end{figure}

As an example we show an implementation of these methods in Figure \ref{fig:tau and chi2}. For this figure we implement the above methods for the correlation function modelling of the results from mock galaxy sample M2.
Presented results are representative of the other samples. The top panel shows the integrated autocorrelation time (IAT) as a function of the chain length $N_{iter}$. 
As shown, the IAT or $\tau$ increases and reaches a plateau (a true autocorrelation time) after $N_{burn}$ iterations (shaded area), marking the number of iterations sufficient for the fit to converge. In this case, for the M2 subsample, $N_{burn} = 41,243$.
This number of course varies from sample to sample, as listed in the table \ref{tab:burnout}.

%

\begin{deluxetable}{ccc}
    \tablecaption{Number of iterations $N_{iter}$ in each chain, number of steps necessary to reach convergence of fit $N_{burn}$ for different mock subsamples. \label{tab:burnout}}
    \tablenum{3}
    \tablehead{\colhead{Sample} & \colhead{$N_{iter}$} & \colhead{$N_{burn}$}}
    \startdata
        M1          &   100,000      &   51,230            \\
        M2          &   103,000      &   41,243             \\
        M3          &   180,000      &   88,370       \\
        M4          &   100,000      &   38,578       \\
        M5          &   100,000      &   30,565        \\
        M6          &   38,000       &   13,030         \\ 
    \enddata
    
\end{deluxetable}

\section{Results}
\label{sec:5_results}
We computed the projected two-point correlation function $w_p(r_p)$ for six volume-limited stellar-mass galaxy samples selected from mock galaxy catalogues populated in the BolshoiP N-body simulation (see section \ref{sec:2_cosm_sim} for the description of the subsample selection). 

\begin{figure*}
    \centering
    \includegraphics{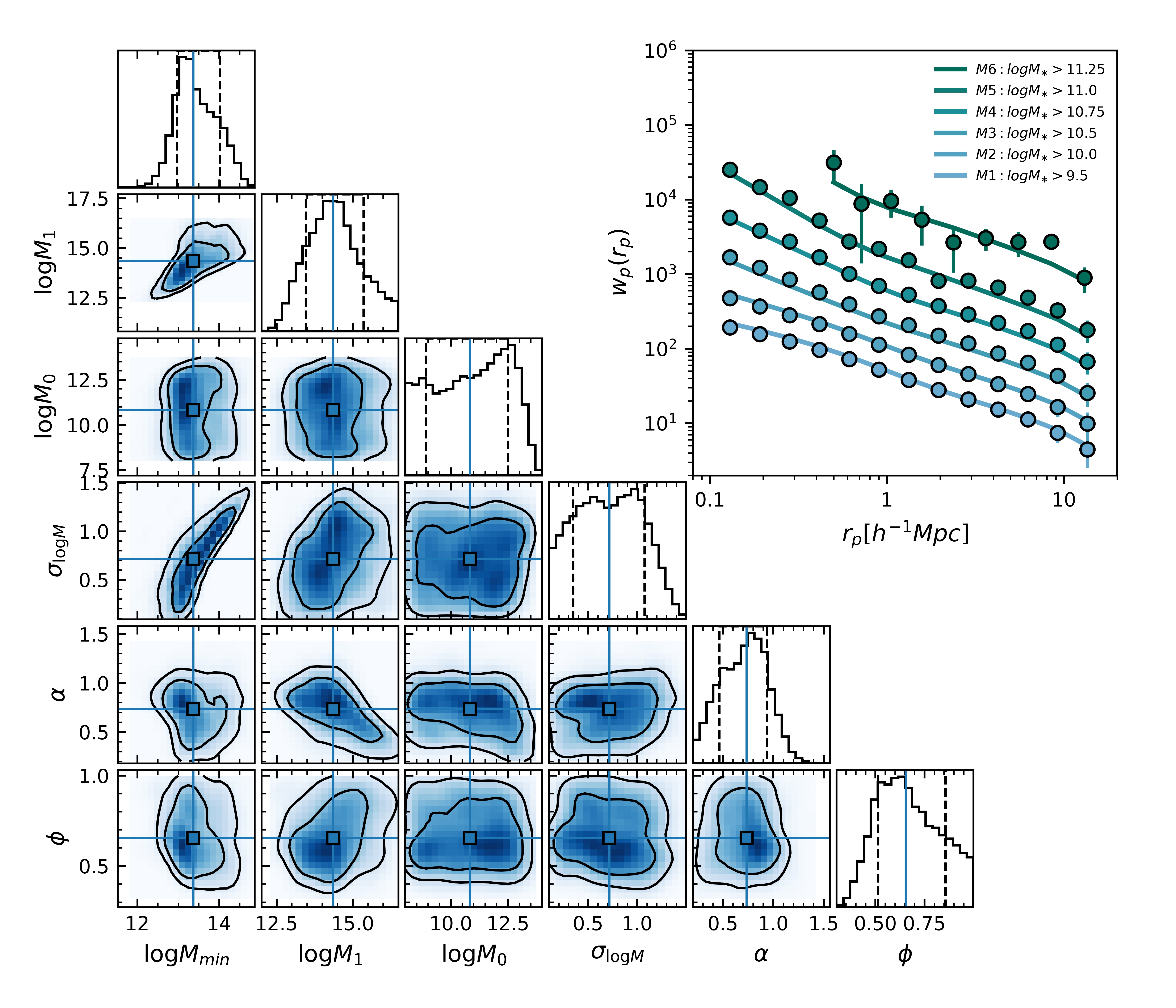}
    \caption{Corner plot: The result of the MCMC fitting of the 6-parameter model to the projected correlation function of the M5 mock galaxy sample. We show only this one for brevity, but each of the correlation function models shown in the upper right corner has a corresponding corner plot similar to this one. 
    The off-diagonal plots show the density maps for a given set of model parameters. The contours represent regions containing 68.3$\%$, 95.5$\%$ of the posterior density. The histograms on the diagonal show the probability distribution functions (PDFs) for the six fitting parameters. The best fit parameters are indicated by blue solid lines, while the dashed lines show the 16th and 84th percentiles for each parameter. Upper right: Projected two-point correlation function $w_p(r_p)$ (filled circles) with the best-fitting 6-parameter halo model (solid lines) for volume limited stellar mass selected mock galaxy subsamples populated in the BolshoiP  simulations. For clarity, both the data points and the best-fitting curves have been shifted by 0.3 dex. }
    \label{fig:corner_plot_sim}
\end{figure*}

For each correlation function measurement, we performed model fitting using the modified model with the DM halo asymmetry parameter included (see section \ref{sec:modified_dm_profile}). 
All measured correlation functions with the best fit are shown in the top right corner of Figure \ref{fig:corner_plot_sim}, while the obtained best-fit parameters are listed in Table \ref{tab:sm_HOD_parameters}.
%

\begin{deluxetable*}{cccccccc}
    \tablenum{4}
    \tablecaption{Best-fit parameters for six volume limited stellar mass selected mock galaxy samples. $M1$ to $M6$  denote different stellar mass subsamples as in Table \ref{tab:subsamples_sm}. All DM halo masses are given in $M_{\odot}$. \label{tab:sm_HOD_parameters}}

    \tablehead{\colhead{Sample} & \colhead{$\log(M_{min})$} & \colhead{$\log(M_1)$} & \colhead{$\log(M_0)$} & \colhead{$\sigma_{logM}$} & \colhead{$\alpha$} & \colhead{$\phi$} & \colhead{$\chi^2/dof$}}

    \startdata
    M1    &  11.11$^{+0.23}_{-0.23}$ &12.56$^{+0.22}_{-0.21}$ &9.55$^{+1.06}_{-1.10}$ &0.46$^{+0.25}_{-0.32}$ &1.10$^{+0.04}_{-0.05}$ &0.85$^{+0.09}_{-0.09}$ & 1.0 \\
    M2    &  11.47$^{+0.22}_{-0.20}$ &12.79$^{+0.24}_{-0.21}$ &9.75$^{+1.20}_{-1.18}$ &0.43$^{+0.23}_{-0.28}$ &1.07$^{+0.05}_{-0.05}$ &0.84$^{+0.09}_{-0.10}$ & 0.9 \\
    M3    &  12.35$^{+0.60}_{-0.78}$ &13.22$^{+0.68}_{-0.43}$ &10.38$^{+1.65}_{-1.57}$ &0.87$^{+0.53}_{-0.36}$ &0.97$^{+0.09}_{-0.10}$ &0.77$^{+0.15}_{-0.14}$  & 0.9 \\
    M4    & 12.40$^{+0.46}_{-0.74}$ &13.86$^{+0.74}_{-0.59}$ &10.58$^{+1.75}_{-1.53}$ &0.75$^{+0.45}_{-0.37}$ &0.93$^{+0.08}_{-0.10}$ &0.69$^{+0.11}_{-0.15}$  & 0.76 \\
    M5    &  13.36$^{+0.39}_{-0.65}$ &14.36$^{+0.90}_{-1.00}$ &10.83$^{+1.92}_{-1.67}$ &0.71$^{+0.37}_{-0.36}$ &0.73$^{+0.27}_{-0.20}$ &0.65$^{+0.14}_{-0.20}$ &  0.53 \\
    M6    & 13.26$^{+0.32}_{-0.32}$ &14.56$^{+0.44}_{-0.46}$ &10.85$^{+1.58}_{-1.68}$ &0.73$^{+0.24}_{-0.27}$ &1.22$^{+0.47}_{-0.48}$ &0.59$^{+0.21}_{-0.26}$ & 0.62
    \enddata
    \tablecomments{The number of degrees of freedom is equal 8 for samples from M1 to M5 (13 measured $w_p$ values plus the number density $n_g$ minus six fitted model parameters), for the most massive sample M6 $dof = 4$ due to a smaller number of correlation function bins (9 measured $w_p$ values).}
\end{deluxetable*}

In Figure \ref{fig:corner_plot_sim} we show a corner plot with the result of the MCMC fitting of the 6-parameter model to the projected correlation function of mock galaxy sample M5 (as a representative of the other results).
Overall, all of the HOD parameters are slightly correlated, with the strongest correlation between $\sigma_{\log M}$ and satellite mass $M_1$. The correlations with parameter $\phi$ are not significantly different from those between other HOD parameters. Parameter $\phi$ is the most strongly correlated with the satellite halo mass $M_1$, which is expected since these two are related to the one-halo term. 
%

\subsection{Model reliability}
\label{sec:average_halo_masses}
In this section we show that the model introduced in this paper provides accurate measurements of the characteristic DM halo masses and halo shape. 

First, the modelled $w_p(r_p)$ can reproduce the shape of the measured correlation function very well, as shown in the top right corner of Figure \ref{fig:corner_plot_sim}. The model proves to be suitable for typical correlation function measurements covering distances $r_p$ from 0.1 to 20 h$^{-1}$ Mpc.

\begin{figure}
    \centering
    \includegraphics{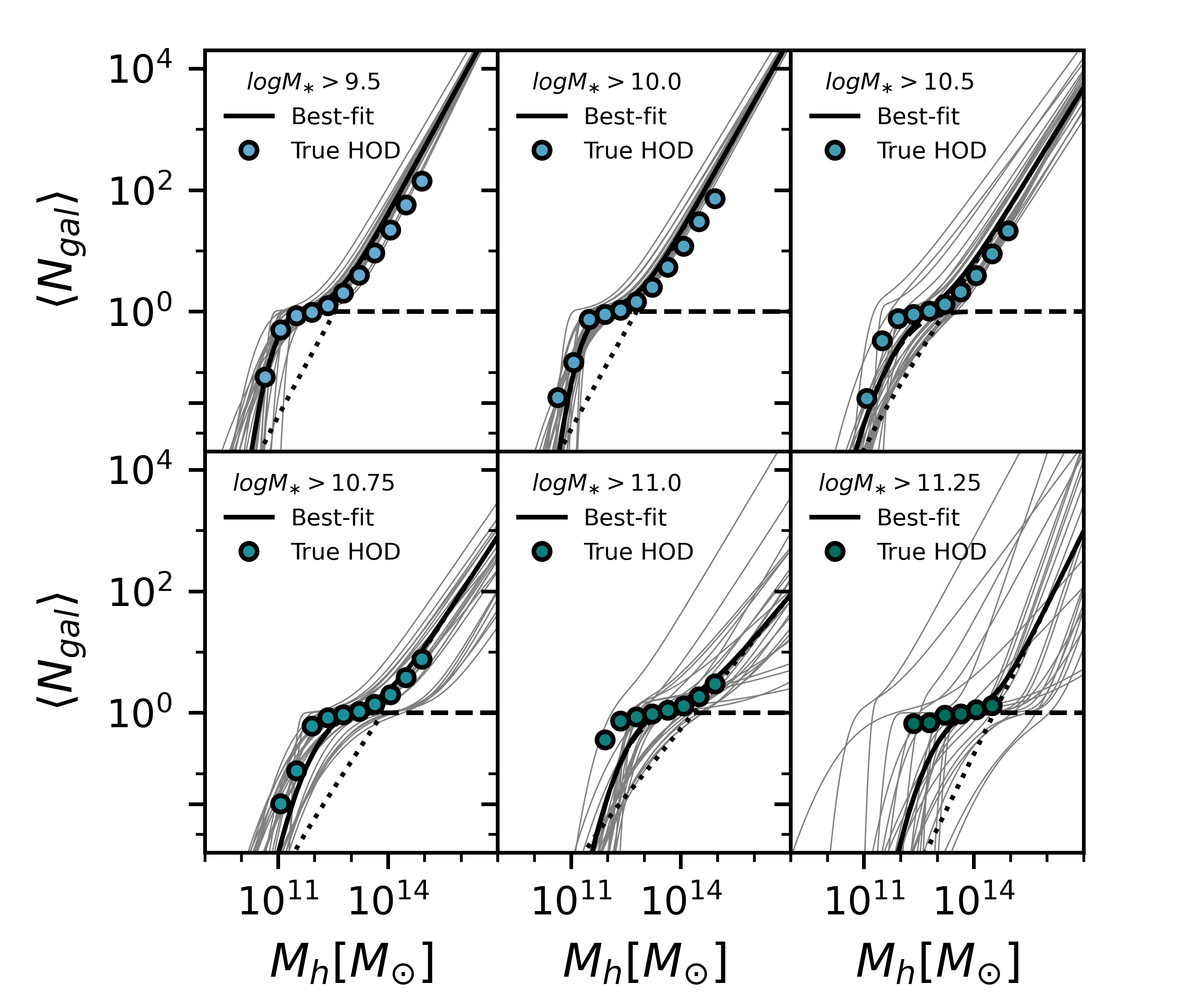}
    \caption{Comparison of the best-fit HODs (black lines) for different mock subsamples with the ``true'' HODs (circles). In each plot, the solid black lines represent the average number of galaxies $\langle N_g \rangle$, the dotted line represents the average number of central galaxies $\langle N_c \rangle$, and the dashed line represents the average number of central galaxies $\langle N_s \rangle$. The series of grey lines in each plot represent 50 randomly selected HODs from the MCMC chains that are within $\Delta \chi^2 < 1$ relative to the best-fitting model.  }
    \label{fig:hod_all_plot}
\end{figure}

Second, the inferred HODs are similar to the ``true'' HODs obtained from the simulations, as is illustrated in Figure \ref{fig:hod_all_plot}, where we plot $\langle N_g \rangle (M)$ for all the samples selected from the mock stellar masses compared to the results obtained from the fit. The largest differences are seen for samples M5 and M6. This may be related to the fact that these samples are the least numerous, making the measurement of the correlation function and the fitting of the model less reliable. 
%

\begin{figure}
    \centering
    \includegraphics[scale=0.9]{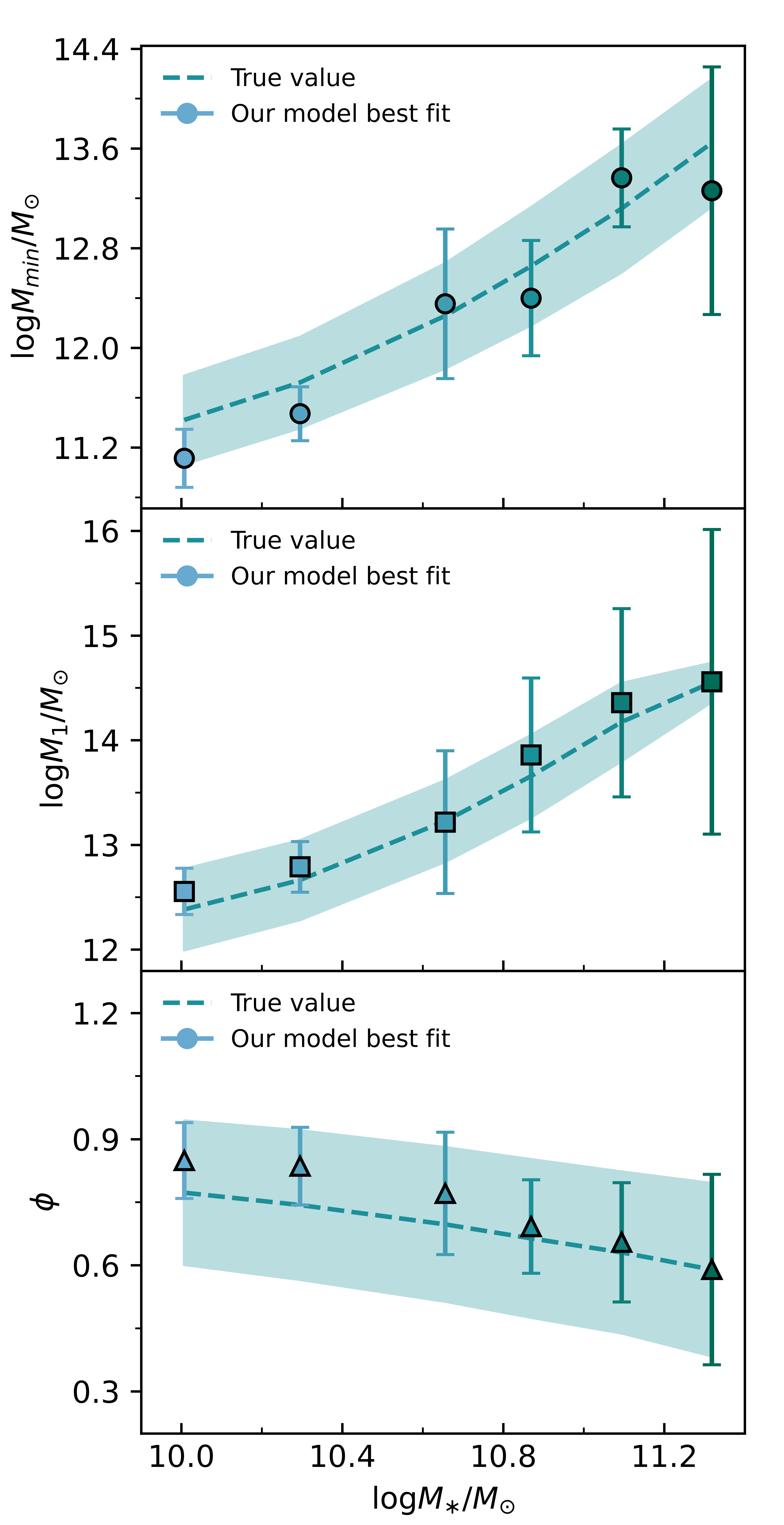}
    \caption{Comparison of the characteristic host halo masses $\log M_{min}$ (top panel) and $\log M_1$ (middle panel) and the asymmetry parameter $\phi$ (bottom panel) obtained directly from the mock catalogue (dashed line) and using our best-fit 6-parameter model (solid circles for $M_{min}$ and solid squares for $M_1$ and triangles for $\phi$). In all figures, the shaded area represents the standard deviation from the mean value obtained from the mock catalogues.}
    \label{fig:best_fit_asym_params}
\end{figure}

Also, as shown in Figure \ref{fig:best_fit_asym_params}, the characteristic host halo masses $M_{min}$ and $M_1$ obtained from the HOD model fit are similar (within $1\sigma$) to the true values for all stellar mass selected galaxy samples. The largest differences are seen for the low-mass galaxy samples, where the $M_{min}$ masses are underestimated. This could be related to the fact that the one-halo term is weakest in these subsamples, which reduces the accuracy of the fit on small scales, affecting the halo mass estimate. Similarly, the underestimation of $M_{min}$ for the most massive galaxy sample may be related to the relatively small number of galaxies affecting the reliability of the correlation function measurement. 

Finally, the most important aspect of the introduction of our model is to provide accurate information about the shape of DM halos. Our model successfully describes the average asymmetry of the host halo. In the bottom panel of Figure \ref{fig:best_fit_asym_params} we compare the best-fit parameter for the DM halo asymmetry $\phi$ (filled symbols) with the ``real'' values measured directly from the BolshoiP mock catalogues (dashed line).
The results are in very good agreement for all stellar mass subsamples - discrepancies between the best-fit parameter $\phi$ and a corresponding ``true'' value are in the 1$\sigma$ range.
%

\subsection{Application to SDSS clustering results from \cite{Zehavi2011} }
\label{sec:zehavi_results}
In their paper \cite{Zehavi2011} present the luminosity and colour dependence of galaxy clustering as seen in the Sloan Digital Sky Survey (SDSS). They measure the two-point correlation function and quantify it using 5-parameter HOD models including an NFW halo density profile and thereby assuming a spherical symmetry of DM halos. 

We use their $w_p(r_p)$ measurements along with the covariance matrices for luminosity threshold selected samples, kindly provided by the authors of that paper, and repeat the HOD modelling, this time using our proposed model. To ease the comparison we adopt the same cosmology and concentration-halo mass relation as \cite{Zehavi2011}. In this section we present the results of this fitting and compare our results to the original measurements presented in \cite{Zehavi2011}.

\begin{figure*}
    \centering
    \includegraphics{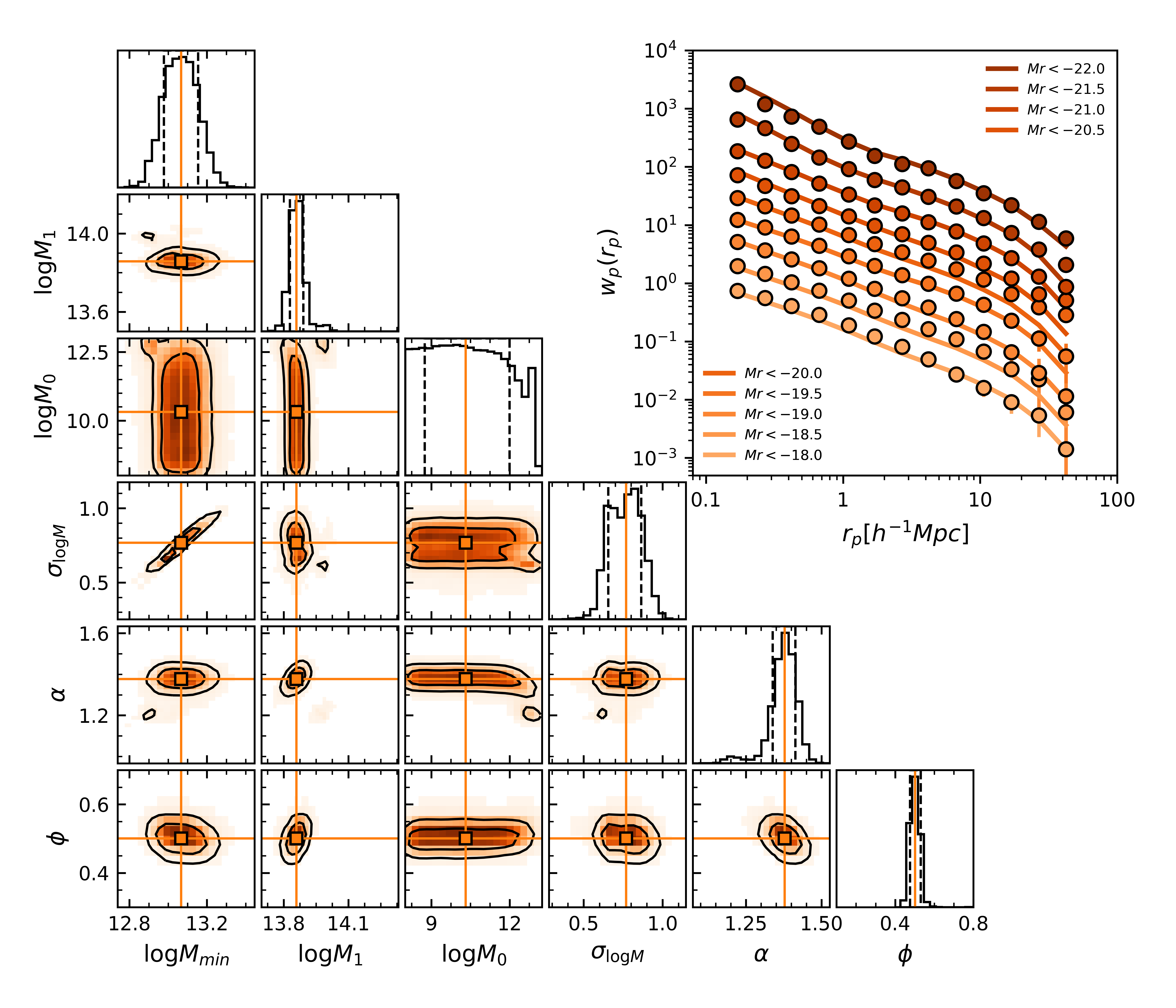}
    \caption{Corner plot: The result of the MCMC fitting of the 6-parameter model to the projected correlation function from \cite{Zehavi2011} sample $M_r^{max} = -21.0$. We show only this one for brevity, but each of the correlation function fits shown in the upper right corner has a corresponding corner plot similar to this one. The off-diagonal plots shows the density maps for given set of model parameters. The contours represent regions containing 68.3$\%$, 95.5$\%$ of the posterior density. Histograms on the diagonal show the probability distribution functions (PDFs) for the six fitted parameters. Best fitting parameters are indicated with orange solid lines, while the dashed lines show the 16th and 84th percentiles for each parameter. Top right: Projected two-point correlation function $w_p(r_p)$ (filled circles) from \cite{Zehavi2011} with the best-fitting 6-parameter models (solid lines) from the SDSS $M_r$ absolute luminosity selected subsamples. For clarity, offsets are applied  to both the data points and the best-fitting curves, i.e., they have been offset by 0.3 dex each. }
    \label{fig:corner_plot_obs}
\end{figure*}

The fitting methods are exactly the same as described in section \ref{sec:fitting}. The correlation function with the best fitting models are shown in the top right panel of Figure \ref{fig:corner_plot_obs}. For each fit, our model is able to reproduce the shape of $w_p(r_p)$. As a representative example of other fits Figure \ref{fig:corner_plot_obs} also shows a corner plot with the results of MCMC fit to the $M_r < -21.0$ projected correlation function from \cite{Zehavi2011}. Parameter $\phi$ is well constrained, and we observe a mild correlation of this parameter with $\log M_1$ and $\alpha$, which is expected as both these parameters are related to the one-halo term.
The parameters of the best fit are shown in the table \ref{tab:Zehavi_params}.
%

\begin{deluxetable*}{cccccccc}
    \tablenum{5}
	\tablecaption{Best-fit parameters obtained in this work for $M_r$ absolute luminosity selected samples from \cite{Zehavi2011}. All DM halo masses are given in $M_{\odot}$. \label{tab:Zehavi_params}}
	\tablehead{\colhead{$M_r^{max}$} & \colhead{$\log(M_{min})$} & \colhead{$\log(M_{1})$} & \colhead{$\log(M_0)$}
                & \colhead{$\sigma_{logM}$} & \colhead{$\alpha$} & \colhead{$\phi$} & \colhead{$\chi^2/dof$}
                }
    \startdata
            -18.0                & 11.77$^{+0.29}_{-0.45}$ &12.96$^{+0.08}_{-0.07}$ &9.93$^{+1.28}_{-1.27}$ &0.84$^{+0.49}_{-0.43}$ &1.13$^{+0.06}_{-0.06}$ &1.28$^{+0.33}_{-0.44}$ & 0.74 \\
            -18.5                &  11.80$^{+0.16}_{-0.34}$ &13.17$^{+0.06}_{-0.06}$ &9.95$^{+1.35}_{-1.31}$ &0.62$^{+0.37}_{-0.41}$ &1.18$^{+0.05}_{-0.04}$ &1.25$^{+0.26}_{-0.41}$ &  0.90 \\
            -19.0                &  11.75$^{+0.09}_{-0.24}$ &13.15$^{+0.07}_{-0.06}$ &9.88$^{+1.27}_{-1.28}$ &0.47$^{+0.26}_{-0.37}$ &1.16$^{+0.03}_{-0.04}$ &1.27$^{+0.29}_{-0.41}$ & 0.83 \\
            -19.5                &  11.97$^{+0.07}_{-0.19}$ &13.33$^{+0.04}_{-0.04}$ &10.22$^{+1.51}_{-1.32}$ &0.40$^{+0.24}_{-0.34}$ &1.27$^{+0.03}_{-0.03}$ &0.71$^{+0.06}_{-0.10}$ & 1.35 \\
            -20.0                &  11.98$^{+0.04}_{-0.06}$ &13.30$^{+0.04}_{-0.04}$ &10.08$^{+1.29}_{-1.41}$ &0.25$^{+0.12}_{-0.19}$ &1.19$^{+0.03}_{-0.03}$ &0.82$^{+0.12}_{-0.08}$ & 2.30 \\
            -20.5                &  12.31$^{+0.03}_{-0.06}$ &13.59$^{+0.07}_{-0.06}$ &10.18$^{+1.45}_{-1.38}$ &0.23$^{+0.10}_{-0.17}$ &1.20$^{+0.03}_{-0.03}$ &0.82$^{+0.16}_{-0.10}$ & 2.37 \\
            -21.0                &  13.07$^{+0.09}_{-0.09}$ &13.86$^{+0.03}_{-0.03}$ &10.31$^{+1.56}_{-1.68}$ &0.77$^{+0.11}_{-0.10}$ &1.38$^{+0.04}_{-0.04}$ &0.50$^{+0.03}_{-0.03}$ & 1.46 \\
            -21.5                &  13.81$^{+0.14}_{-0.07}$ &14.24$^{+0.06}_{-0.18}$ &11.04$^{+2.08}_{-1.91}$ &0.91$^{+0.13}_{-0.06}$ &1.19$^{+0.22}_{-0.10}$ &0.47$^{+0.04}_{-0.10}$ & 1.27 \\
            -22.0                &  14.38$^{+0.24}_{-0.07}$ &14.66$^{+0.06}_{-0.54}$ &11.68$^{+2.38}_{-1.94}$ &0.81$^{+0.24}_{-0.06}$ &1.38$^{+0.50}_{-0.20}$ &0.36$^{+0.03}_{-0.08}$ & 0.97
		\enddata

		\tablecomments{The number of degrees of freedom is equal to 8 for all samples (13 measured $w_p$ values plus the number density $n_g$ minus six fitted HOD parameters).}

\end{deluxetable*}

%

\begin{figure}
    \centering
    \includegraphics[scale = 0.8]{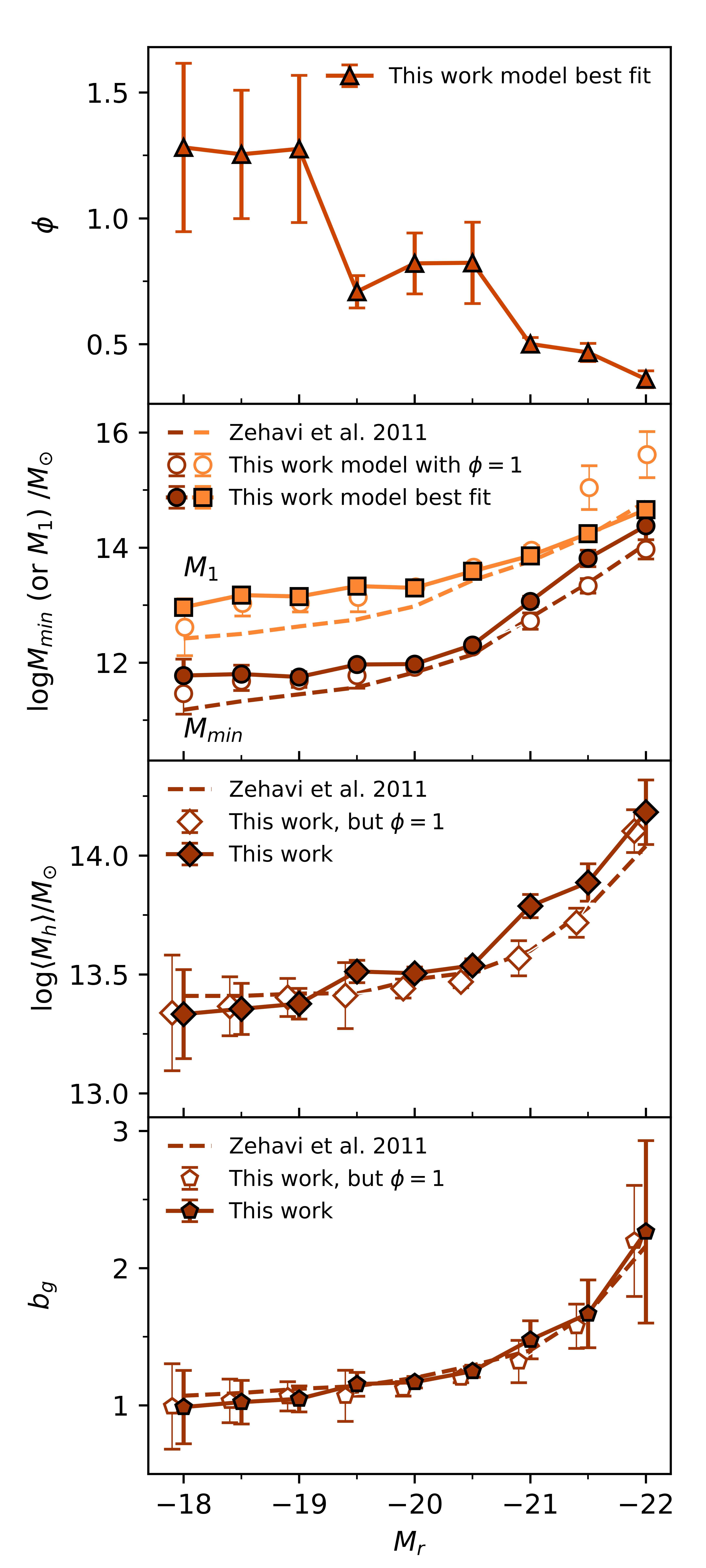}
    \caption{Upper two panels: Comparison of the best-fit host halo asymmetry parameter $\phi$, minimum halo mass $M_{min}$ and satellite halo mass $M_1$ obtained in this work (filled symbols as labelled) and by \cite{Zehavi2011} using the model that assumes spherical symmetry of DM haloes (dashed lines). In case of results from \cite{Zehavi2011} due to the difference in notation we use their $M_1'$. Bottom two panels: Comparison of the average halo masses $\log \langle M_h \rangle$ and galaxy bias $b_g$ estimated using best-fit parameters from this work (filled points) and from \cite{Zehavi2011} (dashed line) as a function of the absolute magnitude $M_r$ of the galaxy sample. }
    \label{fig:best fit zehavi all}
\end{figure}

The most interesting aspect of the 6-parameter halo modelling proposed in this paper is the information about the halo asymmetry. 
In the top panel of Figure \ref{fig:best fit zehavi all} we show how the best-fit host halo asymmetry parameter $\phi$ changes with the luminosity of the galaxy sample.
In general, we observe that the host halo asymmetry becomes stronger with increasing sample luminosity, and hence with increasing host halo mass.
This observation is consistent with our previous conclusions based on mock catalogues: the most massive haloes tend to be the most asymmetric (more prolate). 
However, in the case of measurements based on observations, the change in asymmetry is not continuous. It starts with a plateau of slightly oblate host haloes with $\phi \sim 1.2$ for the $M_r^{max} < -18.0$ to $M_r^{max} < -19.0$ samples. Then for the intermediate samples, from $M_r^{max} < -19.5$ to $M_r^{max} < -20.5$, the asymmetry of the host halo increases up to $\phi \sim 0.8$, finally changing to the most asymmetric (prolate) halos of $\phi$ from 0.5 to 0.3 for the most luminous samples. We also note the changes in the parameter measurement uncertainty, which is largest for the low luminosity subsamples. This could be related to the weak one-halo term observed for these samples. The accuracy of the fit is therefore lower at small scales, which is reflected in a higher uncertainty of the best-fit parameter. 

The value of $\phi$ measured for the sample of galaxies $M_r^{max} < -19.5$ deviates significantly from the observed trend. 
This deviation can be explained by the cosmic variance effect on the measurement caused by the Sloan Great Wall (SGW), a supercluster observed at $z\sim0.8$ \citep[see][]{Gott2005}. According to tests performed by \cite{Zehavi2011}, for the luminosity threshold samples used in our work, the presence of the SGW mainly affects the $M_r^{max} < -20.0$ and $M_r^{max} < -19.5$ samples.

Nevertheless, the observed behaviour of the asymmetry parameter is related to the clustering dependence on luminosity observed by \cite{Zehavi2011}. The correlation functions for $M_r^{max} < -18.5$ and $<-19.5$ samples are nearly identical (hence the plateau of nearly identical results). Then, there is an increase in clustering strength when moving to samples with $M_r^{max} < -20.5$ and $M_r^{max} < -21.0$. Finally, they observe a rapid increase in correlation strength going to $M_r^{max} < -21.5$ and $M_r^{max} < -22.0$, which correlates with the rapid change in halo asymmetry that we observe.

Another point of comparison are the two characteristic halo masses - the minimum halo mass $M_{min}$ and the satellite halo mass $M_1$ - obtained from the best fit\footnote{Please note the difference in notation between this paper and \cite{Zehavi2011} - here we call $M_1$ what in \cite{Zehavi2011} is referred as $M_1'$ and compare accordingly.}.
As shown in the second panel of Figure \ref{fig:best fit zehavi all}, these halo masses differ between the two models.
The $M_{min}$ halo masses obtained by \cite{Zehavi2011} are consistently underestimated (on average by $3\%$ in $\log M_{min}$) with respect to the values obtained with our model. 
The differences are larger for low luminosity galaxy samples ($M_r > -19.0$), but these also have larger uncertainties.
Similarly, the $M_1$ halo masses obtained by \cite{Zehavi2011} are smaller than those obtained in our work, but only for samples with $M_r > -20.0$, the differences are also stronger, reaching on average $4.6\%$ (in $\log M_1$) for these samples. 
For galaxies brighter than $M_r = -20.0$, the values of the satellite halo mass $M_1$ obtained by the two models are almost identical. 
This indicates that the satellite masses are not affected by the halo asymmetry in this luminosity range. 

Although we use the same cosmology and concentration-halo mass relation as \cite{Zehavi2011}, the differences in $M_{min}$ and $M_1$ described in the previous paragraph cannot be related to the halo asymmetry alone. There are other subtle differences between model components that have been shown to play a role in halo mass estimates. The most important are: the halo mass function, the large scale halo bias, and halo exclusion method. In addition, even if we use the same model for the concentration-mass relation our virial mass definition is modified by the parameter $\phi$ (see Appendix \ref{app:model_description}), while in \cite{Zehavi2011} it is not. 
We therefore proceed with the more direct comparison, using exactly the same model as proposed in this paper but fixing $\phi =1$, which represents spherically symmetric halos. The results of the best fitting parameters of this model are represented in Figure \ref{fig:best fit zehavi all} by open symbols. In this case the difference between the model with free asymmetry parameter, and model with fixed  $\phi = 1$ is on average $1\%$ for $\log M_{min}$. However towards the highest luminosity samples ($M_r < -20.5$) this difference increases to $\sim3\%$ and exceeds $1\sigma$ errors.

The situation is similar for satellite halo masses $M_1$. On average the difference is $2\%$ for $\log M_1$. However noticeably for the two most luminous samples it reaches $6\%$. This indicates the halo shape has the strongest influence on the halo mass estimates of the most massive halos which host the brightest galaxies.

In the case of the average host halo masses $\langle M_h \rangle$, obtained using the best fit parameters, the results are very similar for both models, as shown in the third panel of Figure \ref{fig:best fit zehavi all}.
However, we tentatively observe a trend where for the brightest galaxies $M_r < -21.0$, hosted by the most massive haloes, the estimates of $\langle M_h \rangle$ both from \cite{Zehavi2011} and model with fixed $\phi=1$ are underestimated with respect to asymmetric 6-parameter model, while remaining within $1\sigma$ for low and intermediate luminosity galaxies.
This suggests that for the most massive haloes, which are also the most asymmetric (prolate), the assumption of spherical symmetry may influence the average halo mass estimates. These results need to be confirmed using correlation function measurements based on more numerous samples.

Finally, when comparing the large scale galaxy bias $b_g$, obtained using results from both the ``classical'' and the 6-parameter model proposed in this paper, we see no difference between the models. 
The trend of $b_g$ increasing with the luminosity of the galaxy sample is preserved, and the results from all discussed models are well within $1\sigma$ errors - as shown in the bottom panel of Figure \ref{fig:best fit zehavi all} - indicating that the halo asymmetry does not affect the galaxy bias measurements.  

\section{Discussion}
\label{sec:discussion}
\subsection{Model limitations and realistic usability}
Our model can be described as a conventional HOD model. As such, it suffers from the well-known shortcomings of this type of models. In particular, it assumes that the halo mass is the main driver of the galaxy-halo connection. There are known violations of this assumption, commonly referred to as galaxy assembly bias (or simply assembly bias). Meaning that the galaxy occupation is strongly related to several secondary halo properties other than halo mass \citep[see e.g.,][]{Miyatake2016, Artale2018, Zehavi2018, Hadzhiyska2020, Xu2021, Yuan2021}. \cite{Zentner2014} showed that ignoring the assembly bias in halo occupation modelling leads to significant systematic errors. Especially for extreme populations, such as star-forming or quenched galaxies. 

In our model we address only one of the possible sources of assembly bias - halo asymmetry. 
We aim to answer the question: can we build a simple halo model that accounts for halo asymmetry and is able to make reasonably good predictions about galaxy-halo connections? In particular, can it be used to model $w_p(r_p)$ measured for observational data, and provide information about the mass and asymmetry of the halo? As we show in Section \ref{sec:5_results}, the best-fit parameters from the $w_p(r_p)$ fitting of our model are in good agreement with the ``true'' values from the simulations on which the correlation function measurements are based. We are also able to fit our model to observational data and obtain the average host halo asymmetry for any given galaxy sample. 

It should be noted, however, that our model makes a number of assumptions that may not hold for all galaxy samples. 
The first assumption is in the definition of the halo asymmetry parameter $\phi$ itself - it does not describe the shape of the halo, but rather the deviation from spherical symmetry. At this stage we assume that the two axes of the ellipsoidal halo are the same, and measure how the other axes deviate from this symmetry. This is a good first approximation (allowing us to limit the number of free parameters), and as we show in Section \ref{sec:5_results}, the model performs well and is able to accurately retrieve the ``true'' values from the $w(r_p)$ fit. 

The other known weakness is the use of the NFW density profile as a universal recipe for modelling DM haloes, regardless of their size and mass. 
Many studies point out that the NFW density profile can only reproduce the real mass distribution of DM haloes with a very limited accuracy.
For example, at the scale of single galaxy halos, \cite{Gentile2004} showed that the NFW density profiles are inconsistent with the measured (using velocity curves) dark matter distributions of spiral galaxies, because they do not take into account the central density core that occurs in these galaxies. 
It has also been shown that the deviations from the spherical NFW profile increase when we consider the most massive haloes \citep[e.g.,][]{Klypin2016}. These shortcomings are mitigated in our model by the introduction of an asymmetry dependence.

Despite all these known problems, the NFW density profile, coupled with ``classical'' HOD models, is still widely used \citep[e.g. see recent studies by][]{Gao2022, Lange2022, Linke2022, Yung2022, Qin2022, Zhai2022, Harikane2022, Alonso2023, Petter2023}. The reason for this is its applicability to a wide range of data, especially those of limited size. For these samples, systematic errors related to the assembly bias are negligible compared to other uncertainties related to the sample size. 
The model proposed in this paper can easily be used in these types of studies, complementing the ``standard'' halo mass measurement with information on halo asymmetry. However, it should not be used as a method to associate galaxies with the simulated haloes. 

\subsection{Comparison with different models}
With all these limitations in mind, we examine how our modified 6-parameter model compares with other models. We use exactly the same measurements of the correlation functions for stellar mass selected mock galaxy samples (see section \ref{sec:sample_creation}) and fit two additional models. The first model, henceforth called {\it classic}HOD, is virtually identical to the model proposed in this paper. All model components (e.g., concentration-mass relation, halo mass function) are the same, but we fix the parameter $\phi = 1$. With the second model, hereafter called concentrationHOD, we test the influence of the concentration-mass relation and the asymmetry on the modelled correlation function. Again we fix $\phi=1$ and keep the other components the same, except for the halo concentration-mass relation, which we change from the power-law relation (see Appendix \ref{app:model_description}) to one proposed by \cite{Ludlow2016}.

\begin{figure*}
    \centering
    \includegraphics[scale = 0.93]{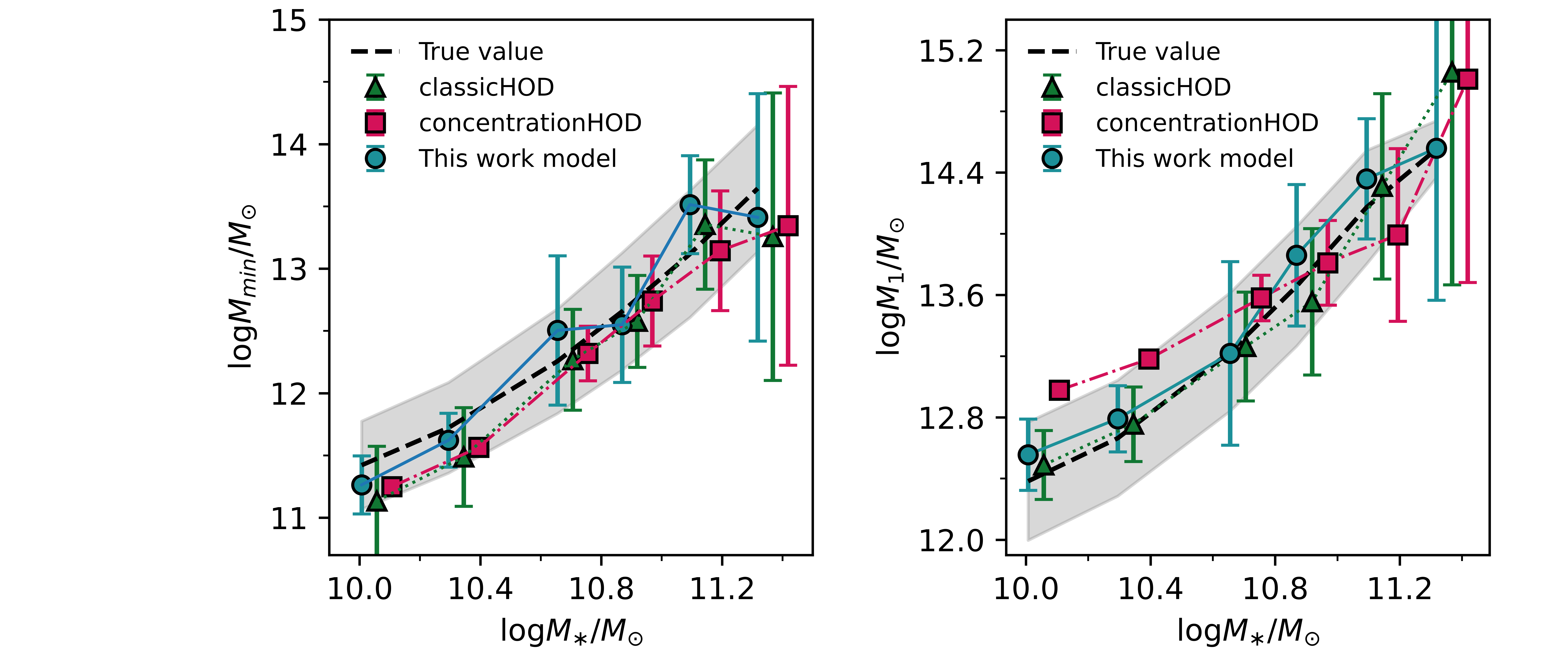}
    \caption{Comparison of best-fit characteristic halo masses $M_{min}$ and $M_1$ obtained via a 6-parameter model fit (solid line), standard HOD model with NFW density profile (dotted line), and HOD model with halo concentration-mass relation from \cite{Ludlow2016} (dotted line). In both figures, the shaded area represents the 1$\sigma$ deviation from the mean true value obtained from the mock catalogues. }
    \label{fig:model_comparison}
\end{figure*}

In Figure \ref{fig:model_comparison} we show the comparison of the best-fit characteristic halo masses $M_{min}$ and $M_1$ obtained from these three models (for mock samples M1-M6). In both plots, the grey area represents the 1$\sigma$ deviation from the mean true value obtained from the mock galaxy catalogues, and different points represent the best-fit results from three models as labelled. As shown in the left panel of Figure \ref{fig:model_comparison}, we do not observe any significant differences between the best-fit $M_{min}$ values, which are consistent within the uncertainties. However, we note that the {\it classic}HOD minimum mass estimates are typically lower than our 6-parameter model results. On average these differences fluctuate around 2$\%$ for $\log M_{min}$.

Similarly, in the right panel of Figure \ref{fig:model_comparison}, the best-fit $M_1$ results from {\it classic}HOD and our 6-parameter model are in very good agreement and are well within $1\sigma$ of the true values. Notably, however, for low mass samples ($\log M_{\ast}< 10.5$) the best-fit $M_1$ values from the {\it concentration}HOD model are higher than the true values, but are in agreement for higher mass samples. This result is another proof that the concentration-mass relation plays a significant role in correlation function models \citep{Artale2018, Zehavi2018, Bose2019, Hadzhiyska2020}. What is important in the context of our work is that the concentration-mass relation has a stronger influence on the satellite halo mass than the halo asymmetry, especially for low mas halos. 

However at the fundamental level of usability, our model does not deviate significantly from the {\it classic}HOD and at the same time provides information about the shape of the halo.

\subsection{DM halo asymmetry and its stellar mass dependence}
In Section \ref{sec:3_DMhalo_shape} we have shown that the shape of the DM halo correlates with the stellar mass of the galaxy.
The majority ($\sim$67$\%$) of galaxies with stellar masses $\log(M/M_{\odot})>11.25$ tend to occupy DM halos of prolate shape (see Figure \ref{fig:T_lum_sm_both}).
Using our modified halo model we are able to reproduce the same results. 
Based on the values of $\phi$ obtained by fitting the model, we observe that the asymmetry of the host halo varies with the stellar mass of the galaxy.
The parameter $\phi$ decreases from $0.85\pm0.10$ for the least massive galaxies to $0.59\pm0.21$ for the most massive.
This means that the average DM halo shape changes from almost spherically symmetric (values of $\phi$ close to 1) to increasingly asymmetric with increasing stellar mass, just as shown in section \ref{sec:3_DMhalo_shape}.

This observation is in broad agreement with previous studies of DM halo shape.
\cite{Allgood2006} examined the dependence of the shape parameters on halo mass and radius in $\Lambda$CDM N-body simulations over the redshift range $z=0-3$.
They found that the majority of halos are prolate at all redshifts, with the fraction of halos that are prolate increasing for halos more massive than the characteristic mass $M_{\ast}$ at a given redshift (in the simulations used by \cite{Allgood2006} $M_{\ast}$ for $z=0$ is $8.0 \times 10^{12}$ M$_{\odot}$).

Similarly, \cite{Hahn2007a,Hahn2007b}, who studied the environmental dependence of the DM halo shape, found that the environment influences the halo shape.
Halos of masses below $M_{\ast}$ tend to be more oblate, and above this mass - favour the prolate shape.
These studies are extended by \cite{Despali2014}, who found, based on three cosmological simulations, that DM halos tend to be prolate regardless of redshift (with a slight tendency to become more triaxial in earlier epochs).
As for the correlation between halo mass and its asymmetry, in the same paper they found that the more massive haloes are less spherical, regardless of the cosmic epoch.

From an observational point of view, massive galaxies tend to cluster more, i.e. they occupy denser environments. 
In terms of models, in particular the HOD model, it is usually interpreted that their DM halo mass is strongly dependent on the galaxy properties, with more luminous and massive galaxies occupying more massive halos \citep[see e.g, ][]{Norberg2002, Abbas2006, Abbas2007, Pollo2006, delaTorre2007, Coil2008, Meneux2008, Abbas2010, Hartley2010, Zehavi2011, Coupon2012, Mostek2013, Marulli2013, Beutler2013, Guo2015, Skibba2015, Durkalec2018, Paul2019}. 
Considering that the most massive haloes are the most asymmetric, we conclude that the DM halo shape must be taken into account when modelling the galaxy correlation function, especially for galaxies with high luminosity and stellar mass.

\section{Summary and Conclusions} 
\label{sec:6_summary}
In this paper we present a 6-parameter model designed to account for halo asymmetry in the modelling of the galaxy's two-point correlation function. The proposed model includes, in addition to the classical 5-parameter HOD, an additional parameter $\phi$ (implemented in the NFW density profile) describing the deviation of the halo shape from spherical symmetry. This parameter is largely related to the more commonly used triaxiality parameter $T$, as it is based on the ratios of the ellipsoid axes (see figure \ref{fig:phi_T}).

In the first part of the paper, we test our model on a sample of mock galaxies populated (using the sub-halo abundance matching method) in BolshoiP N-body simulations.
We measure the real space two-point correlation $w_p(r_p)$ function for six stellar mass selected mock galaxy samples and model these functions with a 6-parameter model including a newly proposed asymmetry parameter $\phi$.
We then compare our best-fit results with the ``real'' values provided by the simulations.

In the second part of the paper we fit our model to the two-point correlation function measurements of \cite{Zehavi2011}, which are based on SDSS observations. 
\cite{Zehavi2011} performed the traditional 5-parameter HOD modelling, which assumes spherical symmetry for DM haloes. We compare their results with those of our model.

The main results and conclusions can be summarised as follows:
\begin{itemize}
    \item We find that (1) the 6-parameter model can reproduce the measured shape of the galaxy correlation function and the halo occupation function quite well, and (2) the new asymmetry parameter $\phi$ and other halo mass parameters computed from the best $w_p(r_p)$ fits are in good agreement (within $1\sigma$ error) with the analogues measured directly from the simulations.

    \item Using best-fit estimates of the asymmetry parameter $\phi$ obtained from modelling the correlation functions of \cite{Zehavi2011} using SDSS data, we show that the halo asymmetry increases with the luminosity of the galaxy samples. The most luminous galaxies $M_r^{max} < -21.0$ are located in the most massive and asymmetric (prolate) halos. The intermediately luminous galaxies $-19.0 > M_r^{max} > -20.5$ reside in halos that are almost spherical (but still prolate), while the least luminous galaxies $M_r > -19.0$ reside in halos that are slightly oblate. In none of the samples is the halo shape perfectly symmetrical.

    \item Comparison of the best-fit characteristic halo masses $M_{min}$ and $M_1$ from \cite{Zehavi2011} (assuming spherical symmetry of the halo) and from this work (including halo asymmetry as a free parameter) shows a $3\%$ difference between the minimum halo masses $\log M_{min}$ for all galaxy samples, with values from \cite{Zehavi2011} being consistently lower. On the other hand, satellite masses $\log M_1$ differ by $4.6\%$ for low and intermediate luminosity samples, while they are in good agreement for bright samples. In case of the symmetrical model assuming $\phi = 1$, the characteristic masses are comparable except for the most luminous samples where the difference reaches $6\%$ in the case of $\log M_1$, and is well above the $1\sigma$ level.

    \item Comparison of the estimates of the mean halo masses $\langle M_h \rangle$ shows agreement between the models for low and intermediate luminosity samples in the SDSS data. For the brightest galaxies, the $\langle M_h \rangle$ estimated using the halo sphericity assumption are lower than those obtained in our work. These differences suggest that for massive haloes the estimates of their mass are sensitive to the assumption of their shape.

    \item Lastly we present that the galaxy bias $b_g$ for the SDSS data is not influenced by the shape of the DM haloes.
\end{itemize}

Overall, the modifications to the halo model proposed in this paper allow us to complement measurements of traditional halo parameters with information about halo asymmetry. 
We have shown that this modified model can reproduce both simulated results and observational measurements of the correlation function very well. Thus, this model can serve as an alternative model when the extended multi-parameter HOD models prove too complex for a given data sample. At the fundamental level, our model performs similarly to ``classical'' HOD models, but additionally provides information about the shape of the halo.

Future developments of this model may include: (i) the introduction of a more complex 2-parameter description of the halo asymmetry, which will allow for more precise discrimination between oblate, prolate and triaxial halos, and, (ii) the inclusion of different and more detailed mass functions, bias and concentration-mass dependencies. Our model as it stands can also be used in studies of the dependence of halo asymmetry on redshift.

We would like to thank the anonymous referee for the useful comments and suggestions.
The authors would also like to thank I. Zehavi for providing their $w_p(r_p)$ and covariance matrix measurements from \cite{Zehavi2011}.
We would like to thank A. Hearin for comments that helped to improve this work.
AD is supported by the Polish National Science Centre grant UMO-2015/17/D/ST9/02121. This work is also supported by Polish National Science Centre grant UMO-2018/30/M/ST9/00757 and Polish Ministry of Science and Higher Education grant DIR/WK/2018/12.
The MultiDark Database used in this paper and the web application providing online access to it were constructed as part of the activities of the German Astrophysical Virtual Observatory as result of a collaboration between the Leibniz-Institute for Astrophysics Potsdam (AIP) and the Spanish MultiDark Consolider Project CSD2009-00064. The Bolshoi and MultiDark simulations were run on the NASA’s Pleiades supercomputer at the NASA Ames Research Center. The MultiDark-Planck (MDPL) and the BigMD simulation suite have been performed in the Supermuc supercomputer at LRZ using time granted by PRACE.
This research made use of \textsc{HaloMod}\footnote{https://pypi.org/project/halomod/} \citep{Murray2021} and \textsc{Emcee} \footnote{https://emcee.readthedocs.io/en/stable/} \citep{Foreman2013}.
This work was completed in part with resources provided by the Świerk Computing Centre at the National Centre for Nuclear Research.

\appendix
\section{Full model description}
\label{app:model_description}
The analytical model of the two-point correlation function proposed in this paper is embedded in the HOD formalism, and as such comes with a set of assumptions. 
Namely, that the galaxies form and evolve in so-called dark matter halos and that the probability $P(N|M)$ of finding a $N$ number of galaxies of a given type residing in the halo is related to its mass $M$. 
For a general description of the HOD models we refer the reader to review papers \cite{Cooray2002} and \cite{Asgari2023}. In our model we also rely on \cite{Jing2002}, \cite{Smith2005}, and \cite{Tinker2005}.

\subsection{Density profile for ellipsoidal halos}
\label{apsec:density_model}
We adapt the common interpretation of a halo as an object having a volume with an averaged overdensity above the critical value $\Delta_{vir}\rho_{crit}(z)$ with respect to the background density. To describe an asymmetrical dark matter density profile of halos we then follow:
\begin{equation}
    \frac{\rho(R)}{\rho_{crit}(z)} = \frac{\delta_{c}}{\frac{R}{R_s}\left(1 + \frac{R}{R_s}\right)^2}
\end{equation}
where $\rho_{crit}(z)$ is a critical density of the universe at given redshift, R is a 3-dimensional vector described by the three ellipsoidal axes, modified to include the asymmetry parameter $\phi$ (as described in Section \ref{sec:modified_dm_profile}, see equations \ref{eq:R} to \ref{eq:R_final}), and finally $\delta_{c}$ is a characteristic overdensity of a halo. Since the ellipsoidal shape of halos in our model is only assumed to be prolate, oblate, or spherical, with axes $a=b$ or $b=c$ (see section \ref{sec:modified_dm_profile}), we define a concentration parameter $c_e \equiv R_{vir}/R_s$, so that the characteristic overdensity is the same as for the NFW profile: 
\begin{equation}
    \delta_{c} = \frac{\Delta_{vir}}{3}\Omega_m(z)\frac{c_e^3}{\ln(1+c_e) - \frac{c_e}{1+c_e}}
\end{equation}
This choice is motivated by the results of \cite{Sheth2001}, who show that ellipsoidal collapse stops at the same density as a spherical collapse \citep[see also][]{Corless2007}. In our work we use two different values of $\Delta_{vir}$ and two different approximations of the concentration-mass relation, depending on the data set to which we apply our model. For simulated mock samples we use $\Delta_{vir}=360$ - the value found for BolshoiP simulations -  and power law $c(M)$ with parameters as described in \cite{Klypin2011}, equation 10 therein. For observational data we use $\Delta_{vir} = 200$ and the concentration function from \cite{Bullock2001} with parameters as described in section 2.3 in \cite{Zehavi2011}. These choices ensure the consistency when comparing the results from different works.

Finally, the virial mass with the asymmetry parameter defined in this paper is:
\begin{equation}
    M_{vir} = \frac{4\pi}{3}\Delta_{vir}\phi^2 R_{vir}^3 \rho_{crit}(z).
    \label{eq:virial_mass}
\end{equation}

\subsection{Correlation function modelling}
The correlation function within the halo occupation framework can be split into two components. The one-halo term $\xi^{1h}(r)$ dominates on scales smaller than the size of a halo (typically  $<1-2$ h$^{-1}$Mpc), the two-halo term $\xi^{2h}(r)$ dominates on larger scales.  Consequently, the correlation function is described as:
\begin{equation}
    \xi(r) = 1 + \xi^{1h}(r) + \xi^{2h}(r).
\end{equation}
\subsubsection{One halo term}
The one halo term depends on the number of galaxy pairs per halo $\langle N(N-1)\rangle$. The halo occupation $\langle N|M\rangle$ is expressed by the sum of central and satellite galaxies (see Eq. \ref{eq:HOD}). The central-satellite pairs can be expressed as $\langle N_{cen}N_{sat}\rangle (M) = N_{cen}(M)N_{sat}(M)$ and the satellite-satellite pairs as $\langle N_{sat}(N_{sat}-1)\rangle = N_{sat}^2(M)$. In the same way the one-halo term is split into central-satellite $\xi_{cs}(r)$ and satellite-satellite $\xi_{ss}(r)$ components:
\begin{equation}
    \xi_{1h}(r) = \xi_{cs}(r) + \xi_{ss}(r),
\end{equation}
The central-satellite component can be calculated in real space using:
\begin{equation}
    1+\xi_{cs} = \frac{1}{\bar{n_g}^2} \int_{M_{vir}(r)}^{M_{high}}dM n(M,z) N_{cen}(M)N_{sat}(M) \frac{\rho(r|M)}{M},
\end{equation}
where ${\bar{n_g}}$ is the mean number density of galaxies; $n(M,z)$ is the halo mass function for which in our model we use the parametrisation proposed by \cite{Tinker2008}; $N_{cen}(M)$ and $N_{sat}(M)$ are the mean number of galaxies within a halo given by Equation \ref{eq:HOD}; $\rho(r|M)$ is the modified NFW density profile described in Section \ref{apsec:density_model}.
Since the satellite-satellite component consists of auto-convolution of the halo density profile, it is easier to compute it in Fourier space:
\begin{equation}
    P_{ss}(k) = \frac{1}{\bar{n_g}^2} \int_{M_{low}}^{M_{high}}dM n(M,z) N_{sat}^2(M) u(k|M)^2
\end{equation}
where $u(k|M)$ is the Fourier transform of the halo density profile $\rho(R)$ calculated numerically. The integrals are from 0 to $\infty$, however, to simplify the calculations we impose realistic halo mass limits of $M_{low} = 10^3$ h$^{-1}$M$_{\odot}$ and $M_{high}=10^{17}$h$^{-1}$M$_{\odot}$.

\subsubsection{Two halo term and halo exclusion}
The two halo term calculations is performed in Fourier space \citep{Tinker2005}:
\begin{equation}
    P_{2h}(k,r) = P_m(k)\frac{1}{n^{\prime2}_g}\int_{M_{low}}^{M_{lim,1}(r)} dM_1 n(M_1,z) N_g(M_1) b_h(r,M_1) u(k|M_1) \int_{M_{low}}^{M_{lim,2}(r)} dM_2 n(M_2,z) N_g(M_2) b_h(r,M_2) u(k|M_2),
\end{equation}
where $b_h(r,M)$ is the large scale dark matter halo bias for which we adopt the parameterisation proposed by \cite{Tinker2010}, $P_m(k)$ is the matter power spectrum, and $n^{\prime}_g$ is the restricted number density:
\begin{equation}
    n^{\prime2}_g (r) = \int_{M_{low}}^{M_{lim,1}(r)} dM_1 n(M_1,z) N_g(M_1) \int_{M_{low}}^{M_{lim,2}} dM_2 n(M_2,z) N_g(M_2).
\end{equation}
The upper integration limits, $M_{lim}(r)$, are related to the halo exclusion method \citep[first proposed by][]{Zheng2004} and are computed using $R_{vir}(M_{lim,1})=r-R_{vir}(M_{min})$ and $R_{vir}(M_{lim,2})=r-R_{vir}(M_1)$. These constraints are imposed to ensure that the halos do not overlap in the transition region between the one-halo term and the two-halo term, i.e, at scales $2 < r < 5$h$^{-1}$Mpc. The general idea is that all pairs between halos must be at separations larger than the sum of the virial radii of these halos, $r\geq R_{vir1}+ R_{vir2}$.
Note that we use the method referred by \cite{Tinker2005} as a ``spherical'' halo exclusion (equations B10 and B11 there). However, since in our model the viral masses and radii are already modified by the asymmetry parameter $\phi$ (see \ref{eq:virial_mass}) there is no need to additionally shape halos by the probability of them being flattened to an ellipsoidal shape, as done by \cite{Tinker2005} in their ``ellipsoidal'' halo exclusion model (equation B12 and B13 therein). In our model $R_{vir1}$ and $R_{vir2}$ are already corrected to be ``ellipsoidal'' by the parameter $\phi$.

This model has not been tested for different cosmological models or parameters other than those mentioned in this paper, nor for redshifts higher than $z=0.1$. 
The model is easily adaptable to higher redshifts given the appropriate redshift dependencies in the various components. Furthermore, the model as proposed in this paper can be subject to improvements such as the inclusion of different and more detailed mass functions, bias and concentration parameterisations. Bearing in mind one of the main aims of this paper, as mentioned in the Introduction, that is to keep the model as simple and computationally lightweight as possible. 

\bibliography{asymmetry_papier}
\bibliographystyle{aasjournal}

\end{document}